%% file: ccc.tex
\documentclass{article}[11pt,a4paper]
\usepackage{amsmath}
\usepackage{amsfonts}
\usepackage{amssymb}
\usepackage{amsthm}
\usepackage{mathrsfs}
\usepackage{verbatim}
\usepackage{graphicx}

\newtheorem{theorem}{Theorem}[section]
\newtheorem{lemma}[theorem]{Lemma}

\newtheorem{definition}[theorem]{Definition}
\newtheorem{proposition}[theorem]{Proposition}

\begin{document}

\setcounter{page}{0}
\title{{\bf From Holant To \#CSP And Back: Dichotomy For Holant$^c$ Problems}}
% Dichotomy Theorems for Some Holant Problems}}

\vspace{0.3in}
\author{Jin-Yi Cai\thanks{University of Wisconsin-Madison.
 {\tt jyc@cs.wisc.edu}}
 \and Sangxia Huang \thanks{Shanghai Jiao Tong University. {\tt hsunshine@gmail.com}}
\and Pinyan Lu\thanks{Microsoft Research Asia. {\tt
pinyanl@microsoft.com}}  }

\date{}
\maketitle

\bibliographystyle{plain}

\begin{abstract}
We explore the intricate interdependent relationship among counting problems,
considered from three frameworks for such problems:
 Holant Problems, counting CSP
and weighted $H$-colorings.  We consider these problems for
general complex valued functions that take boolean inputs.   
We show that results from one framework can be used to
derive results in another, and this happens in both directions.
Holographic
 reductions discover an underlying unity, which is only
revealed when these counting problems are investigated in the
complex domain $\mathbb{C}$.
We prove three complexity dichotomy theorems,
leading to a general theorem for  Holant$^c$ problems.
This is the natural class of Holant problems where one can assign
constants 0 or 1. More specifically, given any
signature grid on $G=(V,E)$ over a set ${\mathscr F}$ of
symmetric functions,
we completely classify the complexity to be in P
or \#P-hard, according to ${\mathscr F}$, of
%\[{\rm Holant}^c = \sum_{\sigma: E \rightarrow \{0,1\}}
\[\sum_{\sigma: E \rightarrow \{0,1\}}
\prod_{v\in V} f_v(\sigma\mid_{E(v)}),\]
%where $f_v \in {\mathscr F} \cup \{\Delta_0, \Delta_1\}$
%($\Delta_0$ and $\Delta_1$ are
%the constant 0 and 1 function).
where $f_v \in {\mathscr F} \cup \{\mbox{{\bf 0}, {\bf 1}}\}$
({\bf 0}, {\bf 1} are the unary constant 0, 1 functions).
Not only is holographic reduction the main tool,
but also the final dichotomy can be only naturally stated
in the language of  holographic transformations.
The proof goes through another dichotomy theorem on
%boolean complex weighted \#CSP.\\
boolean complex weighted \#CSP.

%%%%%%% do we need that??? i am omitting it.
%\noindent {\bf ACM Subject Classification:} 
%F.2.0 [Theory of Computation] Analysis of Algorithms and
%Problem Complexity
%
%\noindent {\bf Keywords:} Holant problem, \#CSP, 
%holographic reduction, dichotomy
\end{abstract}

\newpage

\input{intro}

%
\input{preliminaries}
\input{ternary}

\input{holant-to-CSP}

\input{dichotomy}

\input{bib}
\newpage

%\section*{Appendix}

% \input{degenerate}

%\input{symmetric}

\noindent
{\Large \bf Appendix}
\input{propositions}

\input{JYC-list-of-T.tex}

\input{orthogonal.tex}

\end{document}

%% file: intro.tex
\section{Introduction}
In order to study the complexity of counting problems, several
interesting frameworks have been proposed. One is called counting
Constraint Satisfaction Problems
(\#CSP)~\cite{Bulatov06,Bulatov08,BulatovD03,weightedCSP,FederV98}.
Another well studied framework is called $H$-coloring
or Graph Homomorphism,
 which can be viewed as a special case of
\#CSP
problems~\cite{BulatovG04,BulatovG05,Homomorphisms,DyerGP06,acyclic,DyerG00,DyerG04,GJGT,Hell}.
Recently, we proposed a new refined framework called Holant 
Problems~\cite{FOCS08,STOC09} inspired by Valiant's Holographic 
Algorithms~\cite{HA_FOCS,AA_FOCS}.
One reason such frameworks are interesting is because the language
is {\it expressive} enough so that they can express many natural
counting problems, while {\it  specific} enough so that we can prove
\emph{dichotomy theorems} (i.e., every problem in the class
is either in P or \#P-hard)~\cite{CSPBook}.
By a theorem of Ladner, if P $\not =$ NP, or P $\not =$ \#P,
then such a dichotomy for NP or \#P 
is \emph{false}.
Many natural counting problems can be expressed in all three 
 frameworks. This includes 
counting the number of vertex covers,
the number of $k$-colorings in a graph, and many others. However,
{some}
 natural and important counting problems, such as counting the number of
perfect matchings in a graph,  {\it cannot} be expressed as a graph
homomorphism function~\cite{freedman-l-s}, 
but can be naturally expressed as a Holant Problem.
Both \#CSP and Graph Homomorphisms can be viewed as 
special cases of Holant Problems.
The Holant framework of counting problems makes a finer
complexity classification.
% due to its expressibility. 
A rich mathematical structure is uncovered in the 
Holant framework regarding the
complexity of counting problems, which is sometimes difficult even to state
in \#CSP. This is particularly true when we apply holographic
reductions~\cite{HA_FOCS,AA_FOCS,FOCS08}.
%This is also our main language in this paper.

%This notion is motivated by holographic
%reductions proposed by Valiant~\cite{HA_FOCS,AA_FOCS}. 
We give a brief description of the Holant framework here.
%and a more formal definition is given in
%Section~\ref{sec:background}. 
A {\it signature grid} $\Omega = (G,
{\mathscr F}, \pi)$ is a tuple, where $G=(V,E)$ is a graph, and $\pi$
labels each $v \in V$ with a function $f_v \in {\mathscr F}$. We consider
all edge assignments (in this paper 0-1 assignments).
 An assignment $\sigma$ for every $e\in E$
gives an evaluation $\prod_{v \in  V} f_v(\sigma\mid_{E(v)})$, where
$E(v)$ denotes the incident edges of $v$,
and $\sigma\mid_{E(v)}$ denotes the restriction of $\sigma$  to $E(v)$.
 The  counting problem on the instance  $\Omega$
is to compute
\[{\rm Holant}_\Omega=\sum_{\sigma}
\prod_{v\in V} f_v(\sigma\mid_{E(v)}).\] 
For example, consider the {\sc Perfect
Matching} problem on $G$. This problem corresponds to attaching the
{\sc Exact-One} function at every vertex of $G$, and then consider
all 0-1 edge assignments. In this case, ${\rm Holant}_\Omega$ counts
the number of perfect matchings. If we use the {\sc At-Most-One}
function  at every vertex, then we count all (not necessarily
perfect) matchings.
We use the notation
Holant(${\mathscr F}$) to denote the class of Holant problems where all
functions are given by ${\mathscr F}$.

To see that Holant is a more expressive framework, we show that
every \#CSP problem can be directly 
simulated by a Holant problem. Represent
an instance of a \#CSP problem by  a bipartite graph where LHS are
labeled by variables and RHS are labeled by constraints. Now the
signature grid $\Omega$ on this bipartite graph is as follows: Every
variable node on  LHS is labeled with an {\sc Equality} function, every
constraint node on RHS is labeled with the given constraint function. Then
${\rm Holant}_\Omega$ is exactly the answer to the counting CSP
problem. In effect, the {\sc Equality} function on a node in LHS
forces the incident edges to take the same value; this
effectively reduces to a vertex assignment on LHS as
in \#CSP. 
%It follows that \#CSP problems are precisely the special
%case of Holant problems on bipartite graphs where every node in  LHS
%is attached an {\sc Equality} function.
We can show that \#CSP  is equivalent to
Holant problems where  {\sc Equality} functions 
of  $k$ variables, for arbitrary $k$  (denoted by $=_k$), are 
freely  and implicitly available  as constraints.
%
%However the Holant framework is more refined.
%However, there is no obvious way to reverse  the above process.
%%% JYC: i made it stronger. there are fixed degree tractable ones
%%% but hard for CSP. so it is provably not reversible.
However, this process provably cannot be reversed in general (if
P $ \not = $ \#P).
%by simulating Holant with \#CSP in general.
While \#CSP is the same as  adding all $=_k$ to Holant,
 %(actually 3-way
%{\sc Equality} $=_3$ will do), 
the effect of this is non-trivial.
% adding $=_3$ is non-trivial.
From the lens of holographic transformations, $=_3$ is a full-fledged
non-degenerate symmetric function of arity 3.
%Additionally,

Meanwhile, starting from the Holant framework, rather than assuming
{\sc Equality} functions are free,
one can consider new classes of counting problems
which are
difficult to express as \#CSP problems.
One such class, called Holant$^*$ Problems~\cite{STOC09},
is the class of Holant Problems where all 
unary functions are freely available.  If we allow only
two special unary
functions
%functions $[1,0]$ and $[0,1]$
{\bf 0} and {\bf 1} as freely available, then we obtain
the family of counting problems called
 Holant$^c$ Problems,
which is even more appealing.   This is the class
of all Holant Problems (on boolean variables) where one can 
set any particular edge (variable) to 0 or 1 in an input graph.
 
Previously we  proved a dichotomy theorem for
${\rm Holant}^*(\mathscr{F})$, where $\mathscr{F}$
is any set of complex-valued symmetric functions~\cite{STOC09}.
%Holant$^*$ Problems is studied primarily as a technical bridge
%to other classes of counting problems;
It is used to prove a dichotomy theorem for \#CSP in \cite{STOC09}.
For ${\rm Holant}^c(\mathscr{F})$ we could only
prove a dichotomy theorem for real-valued functions.
In this paper we manage to traverse in the other direction, going from \#CSP
to Holant Problems.
%derivations. 
First we establish a dichotomy theorem
for   a special Holant class.
%Problem involving exactly one
%non-degenerate symmetric signature of arity 3,
Second we prove  a more general dichotomy for bipartite Holant Problems.
Finally by going through  \#CSP,
we prove a dichotomy theorem for complex-valued Holant$^c$ Problems.
Now we describe our results in more
detail.

A symmetric function $f: \{0,1\}^k \rightarrow \mathbb{C}$
will be written as $[f_0,f_1,\ldots,f_k]$, where $f_j$ is the value
of $f$ on inputs of Hamming weight $j$.
%(E.g., the constant unary functions {\bf 0} and {\bf 1} are
%$[1,0]$ and $[0,1]$.)
Our first main result (in
Section \ref{sec-ternary}) is a dichotomy theorem for
%ternary functions. That is a dichotomy theorem
${\rm Holant}(\mathscr{F})$, where
$\mathscr{F}$ contains a single ternary function $[x_0, x_1, x_2, x_3]$.
More generally, as
proved by holographic reductions,
 we get a dichotomy theorem for 
${\rm Holant}([y_0, y_1, y_2] | [x_0, x_1, x_2, x_3] )$
defined on 2-3 regular bipartite graphs. 
Here the notation indicates that every vertex of degree 2
on LHS has label $[y_0, y_1, y_2]$ and every vertex of degree 3
on RHS has label $[x_0, x_1, x_2, x_3]$.
This is the foundation
 of the remaining two dichotomy results in this paper.
Previously we proved a dichotomy theorem 
for  ${\rm Holant}([y_0, y_1, y_2] | [x_0, x_1, x_2, x_3] )$, when
all $x_i$, $y_j$ take values in $\{0,1\}$~\cite{FOCS08}.
Kowalczyk extended this to $\{-1,0,1\}$ in \cite{mike}.
In \cite{TAMC}, we gave a dichotomy theorem for 
${\rm Holant}([y_0, y_1, y_2] | [1,0,0,1] )$, where $y_0, y_1, y_3$
%take arbitrary real numbers. Finally
take arbitrary real values. Finally
this last result was  extended to arbitrary complex numbers~\cite{cai}.
Our result here is built upon these results, especially~\cite{cai}.

Our second result (Section \ref{sec-holant-to-csp})
is a dichotomy theorem, under a mild condition,
%%%%
%%%% JYC
%%%%  I feel the results in sec 4 {sec-holant-to-csp}
%%%%  is never fully stated in generality, except in the generic case.
%%%%  would it be easy to state a clean statement includng the
%%%%  the non-generic case?  it would be nice to just cite
%%%% what we obtain in sec 3, and move on. instead of mostly refer to
%%%% \cite{cai} with Kowalczyk-Cai.
%%%%
for bipartite Holant problems
${\rm Holant}( \mathscr{F}_1
| \mathscr{F}_2)$.
%where $\mathscr{F}_1$ (respectively $\mathscr{F}_2$)
%contains  some non-degenerate function $[y_0,y_1, y_2]$ of arity 2
%(respectively some non-degenerate $[x_0,x_1,x_2,x_3]$ of arity 3).
To prove that, we first use holographic reductions
%To prove that, in a ``generic'' case,  we first use holographic reductions
%%% i think trechnically i should say the above, namely only in 
%%% a ``generic'' case, to go to [1 0 0 1]
%%% but i think i will leave it vague.
 %to transform the problem to ${\rm Holant}(\mathscr{F}'_1
to transform it to ${\rm Holant}(\mathscr{F}'_1
| \mathscr{F}'_2)$, where we transform
some non-degenerate function $[x_0,x_1,x_2,x_3] \in \mathscr{F}_2$ 
%is transformed
to the {\sc Equality} function $(=_3) = [1, 0, 0, 1] \in \mathscr{F}'_2$.
Then we prove that we can ``realize'' the binary {\sc Equality} function
 $(=_2) = [1,0,1]$ in the 
left side and reduce the problem to
$\#{\rm CSP}(\mathscr{F}'_1 \cup \mathscr{F}'_2)$.
This is a new proof approach.
Previously in~\cite{STOC09},
we reduced a \#CSP problem to a Holant problem and obtained results for \#CSP.
Here, we go the opposite way, using
  results for \#CSP to prove dichotomy theorems for Holant problems.
This is made possible
by our complete dichotomy theorem for boolean complex
weighted \#CSP~\cite{STOC09}. We note that proving this over
$\mathbb{C}$ is crucial, as holographic reductions naturally
go beyond $\mathbb{R}$. We also note that
our dichotomy
theorem here does not require the functions
in $\mathscr{F}_1$ or $\mathscr{F}_2$ to be symmetric.
This will be useful in the future.
%So the theorem  will be useful for 
%can have future applications
%in dichotomy theorems for more general Holant problems.

Our third main result, also the initial motivation of this
%%% JYC i cut out the word "whole"
work, is a dichotomy theorem for
symmetric complex Holant$^c$ problems. This improves
our previous result in \cite{STOC09}.
% which gives a dichotomy
%theorem for symmetric real Holant$^c$ problems.
We made a conjecture in \cite{STOC09} that the dichotomy theorem
stated as Theorem~\ref{thm:holnat-c} is also true for
symmetric complex functions. It turns out that
this conjecture is not correct as stated. 
For example, 
Holant$^c([1,0,i,0])$ is tractable (according to our new theorem),
but not included in the tractable cases by the conjecture.
After isolating these new tractable cases
we prove {\it everything else} is \#P-hard.
Generally speaking, 
non-trivial and previously unknown tractable cases are what make
dichotomy theorems particularly interesting, but at the same
time make them more difficult to prove (especially for hardness proofs,
which must ``carve out'' exactly what's left.). 
The proof approach here is also different from
that of \cite{STOC09}. In \cite{STOC09}, the idea is to interpolate all
unary functions and then use the results for Holant$^*$ Problems.
Here we first prove that we can realize some non-degenerate
ternary function, for which we can use the result of our first
dichotomy theorem. Then we use our second dichotomy theorem
to further reduce the problem to \#CSP and obtain a dichotomy theorem
for Holant$^c$.

The study of Holant Problems is strongly influenced by the
development of holographic 
algorithms~\cite{HA_FOCS,AA_FOCS,STOC07,FOCS08}.
% and 
%holographic
%reductions~\cite{HA_FOCS,AA_FOCS,STOC07,FOCS08}. Indeed,
Holographic reduction is a primary technique in the proof of these
dichotomies, both for the tractability part and the 
hardness part. More than that---and this seems to be the 
first instance---holographic reduction
%as we can see from the results of this paper,
 even provides the correct language for the {\it
statements} of
these dichotomies. 
%Without using holographic reductions to
%normalize a given set of functions, it is not easy to even fully describe
Without using holographic reductions, it is not easy to even fully describe
what are the tractable cases in the dichotomy theorem.
Another interesting observation is that by employing
holographic reductions, complex numbers appear naturally
and in an essential way.
% Thus
%our functions ${\mathscr F}$ are complex valued.
Even if one is only interested in integer or real valued
counting problems, in the complex domain $\mathbb{C}$
 the picture becomes whole.
%As Jacques Hadamard said:
%``The shortest path between two truths in the real domain
%passes through the complex domain.'' (Jacques Hadamard)
%%%%%%%%%% JYC: do you like th French one?
%%%``Le plus court chemin entre deux v\'{e}rit\'{e}s dans le domaine
%%%r\'{e}el passe par le domaine complexe.'' ---Jacques Hadamard.
``It has been written that the shortest and best way between two 
truths of the real domain often passes through the imaginary one.'' 
---Jacques Hadamard.

%%%%%%%%%%%%

%% file: preliminaries.tex
\section{Preliminaries}\label{sec:background}
%
%\subsection{Problems and Definitions}
Our functions take values in $\mathbb{C}$ by default.
Strictly speaking complexity results should be restricted to
computable numbers in the Turing model; but it is more convenient to
express this over $\mathbb{C}$.
We say a problem is tractable if it is computable in P.
The framework of
 Holant Problems is defined for functions
mapping any  $[q]^k\rightarrow \mathbb{C}$ for a finite $q$.  Our
results in this paper are for the Boolean case $q=2$.
%So we give the following definitions for $q=2$ for notational simplicity,
%and these can be
%easily extended to arbitrary $[q]$.

%As stated, a {\it signature grid} $\Omega = (G, {\mathscr F}, \pi)$
%consists of a graph $G=(V,E)$ with each vertex labeled by
%a function $f_v \in {\mathscr F}$.
Let ${\mathscr F}$ be a set of functions.
A {\it signature grid} $\Omega = (G,
{\mathscr F}, \pi)$ is a tuple, where $G=(V,E)$ is a graph, and $\pi$
labels each $v \in V$ with a function $f_v \in {\mathscr F}$.
 The Holant problem on instance
$\Omega$ is to compute ${\rm Holant}_\Omega=\sum_{\sigma: E \rightarrow
\{0,1\}}
\prod_{v\in V} f_v(\sigma\mid_{E(v)})$, a sum over all 0-1 edge
assignments, of the products of the function evaluations
at each vertex. A function $f_v$ can be represented as a truth table.
It will be more convenient to
denote it as  a tensor in $({\mathbb{C}}^{2})^{\otimes \deg(v)}$,
or a vector in ${\mathbb{C}}^{2^{\deg(v)}}$,
when we perform holographic tranformations. We also call it
a {\it signature}.
We denote by $=_k$ the {\sc Equality} signature of arity $k$.
A symmetric function $f$ on $k$ Boolean variables
can be expressed by $[f_0,f_1,\ldots,f_k]$, where $f_j$ is the value
of $f$ on inputs of Hamming weight $j$.
Thus, for example, ${\bf 0} = [1,0]$, ${\bf 1} = [0,1]$
and $(=_k)=[1,0,\ldots,0,1]$ (with $(k-1)$ 0's).
%A Holant problem is parameterized by a set of signatures ${\mathscr F}$.
\begin{definition}
Given a set of signatures ${\mathscr F}$, we define a counting problem
${\rm Holant}({\mathscr F})$:

Input: A {\it signature grid} $\Omega = (G, {\mathscr F}, \pi)$;

Output: ${\rm Holant}_\Omega$.
\end{definition}

We would like to characterize the complexity of Holant problems in
terms of its signature set ${\mathscr F}$. Some special families of Holant
problems have already been widely studied. For
example, if ${\mathscr F}$ contains all {\sc Equality} signatures
$\{=_1, =_2, =_3,\cdots\}$, then this is exactly the weighted \#CSP
problem. In \cite{STOC09}, we also introduced the following two special families of Holant problems
by assuming some signatures are freely available.

\begin{definition}
Let ${\mathscr U}$ denote the set of all unary signatures. 
Then ${\rm Holant}^* ({\mathscr F}) = 
{\rm Holant}({\mathscr F}\cup {\mathscr U})$.
%Given a set
%of signatures ${\mathscr F}$, we use ${\rm Holant}^* ({\mathscr F})$ to
%denote ${\rm Holant}({\mathscr F}\cup {\mathscr U})$.
\end{definition}

\begin{definition}
Given a set of signatures ${\mathscr F}$, we use
${\rm Holant}^c
({\mathscr F})$ to denote 
${\rm Holant}({\mathscr F}\cup \{ {\bf 0}, {\bf 1}\})$.
\end{definition}

Replacing a signature $f\in {\mathscr F}$ by a constant multiple $cf$, where $c
\neq 0$, does not change the complexity of ${\rm Holant}({\mathscr
F})$. 
%So we  view $f$ and $cf$ as the same signature.
An important property of a signature is whether  it is degenerate.
\begin{definition}
A signature is degenerate iff it is a
%called degenerate iff it can be decomposed into a
tensor product of unary signatures. In particular, a symmetric signature in ${\mathscr F}$
is degenerate iff it can be expressed as
$\lambda[x,y]^{\otimes k}$.
\end{definition}

To introduce the idea of holographic reductions, it is convenient
to consider bipartite graphs.  This is without loss of generality.
For any general graph, we can make it
bipartite by adding an additional vertex on each edge, and giving each
new vertex  the {\sc Equality} function  $=_2$ on 2 inputs.

We use ${\rm Holant}( {\mathscr G}|{\mathscr R})$ to denote all counting
problems, expressed as Holant problems on bipartite graphs
$H=(U,V,E)$, where each signature for a vertex in $U$ or $V$ is from
${\mathscr G}$ or ${\mathscr R}$, respectively.  An input instance
for the bipartite
 Holant problem is a bipartite signature grid and is denoted as $\Omega =
(H, {\mathscr G}|{\mathscr R},\pi)$. Signatures in ${\mathscr G}$ are
denoted by column vectors (or
contravariant tensors); signatures in ${\mathscr R}$ are
denoted by row vectors (or covariant
tensors)~\cite{dodson}.

One can perform (contravariant and covariant) tensor
transformations on the signatures.
% which may
%produce exponential cancelations in tensor spaces.
We will define a simple version of  holographic reductions,
which are invertible.
%They are called holographic because they may 
%produce exponential cancellations in the tensor space.
Suppose  ${\rm Holant}( {\mathscr G}|{\mathscr R})$ and
 ${\rm Holant}( {\mathscr G'}|{\mathscr R'})$ are two Holant problems defined for the same
family of graphs, and
 $T \in {\bf GL}_2({\mathbb C})$ is a basis.
We say that there is an (invertible) holographic
reduction from  ${\rm Holant}( {\mathscr G}|{\mathscr R})$ to
 ${\rm Holant}( {\mathscr G'}|{\mathscr R'})$, if
the {\it contravariant} transformation
$G' = T^{\otimes g} G$ and the {\it covariant} transformation
$R=R' T^{\otimes r}$ map $G\in {\mathscr G}$ to $G'\in {\mathscr G'}$
and $R\in {\mathscr R}$  to $R' \in {\mathscr R'}$, 
%%%% JYC I added the phrase.  tobe birectional
and vice versa,
where $G$ and $R$ have arity $g$ and $r$ respectively.
(Notice the reversal of directions when the
transformation $T^{\otimes n}$ is applied. This is the meaning of
{\it contravariance} and {\it covariance}.)

\begin{theorem}[Valiant's Holant Theorem]\label{thm:holant}
Suppose there is  a holographic reduction from $\#{\mathscr G}|{\mathscr
R}$ to $\#{\mathscr G'}|{\mathscr R'}$ mapping signature grid $\Omega$
to $\Omega'$, then ${\rm Holant}_{\Omega} = {\rm Holant}_{\Omega'}.$
\end{theorem}

In particular,  for invertible  holographic reductions from ${\rm Holant}( {\mathscr G}|{\mathscr R})$ to
 ${\rm Holant}( {\mathscr G'}|{\mathscr R'})$, one problem
is in P iff the other one is in P, and similarly one problem is \#P-hard
iff the other one is also \#P-hard.

In the study of Holant problems, we will often
 transfer between bipartite and
non-bipartite settings. When this does not cause confusion,
 we do not 
distinguish signatures between  column vectors (or
contravariant tensors) and  row vectors (or covariant
tensors).  Whenever we write a transformation as $T^{\otimes n} F$ or $T \mathscr{F}$,
we view the signatures as  column vectors (or
contravariant tensors);
  whenever we write a transformation as $F T^{\otimes n} $ or $\mathscr{F} T $,
we view the signatures as  row vectors (or covariant
tensors).

%% file: ternary.tex
\section{Dichotomy Theorem for Ternary Signatures}\label{sec-ternary}
In this section, we consider the complexity of ${\rm Holant}([x_0,x_1,x_2,x_3])$.
If $[x_0,x_1,x_2,x_3]$ is degenerate, it is trivially tractable,
since a degenerate signature factors as a tensor product and
the signature grid simply decomposes into isolated edges. In the following
we always assume that it is non-degenerate.
Given a non-degenerate signature $[x_0,x_1,x_2,x_3]$,
we can find a non-zero tuple $(a,b,c)$ (unique up to a scalar factor),
%where $a, b$ and $c$ are not all zero,
%%%% JYC: you had "unary ***???*** not all zero $(a,b,c)$ (up to a scale)"
such that $a x_0 +b x_1 + c x_2=0 $ and $a x_1 +b x_2 + c x_3=0 $.
If $c \neq 0$, the sequence $[x_0,x_1,x_2,x_3]$ is a second order linear
recurrence
sequence. Its characteristic equation is $a + b \lambda+ c \lambda^2 =0$.
We can write down an expression for this sequence $[x_0,x_1,x_2,x_3]$ by the
eigenvalues.
%roots of the equations and two coefficients depend on its initial values.
When $c =0$ and $a \neq 0$, we can consider its reverse sequence.
The case $a=c=0$ can be viewed as a limiting case. Statedly succinctly,
the sequence $[x_0,x_1,x_2,x_3]$ can always be expressed by
one of the following three categories (with the convention that
$\alpha^0 =1$, and $k \alpha^{k-1} =0$ if $k=0$, even when $\alpha=0$):
\begin{enumerate}
	\item $x_k=\alpha_1^{3-k}\alpha_2^k+\beta_1^{3-k}\beta_2^k$,
where $\det \left[
        \begin{array}{c c}
                \alpha_1 & \beta_1 \\
                \alpha_2 & \beta_2
        \end{array}
\right] \not = 0$;
	\item $x_k=Ak\alpha^{k-1}+B\alpha^k$, where $A \neq 0$;
	\item $x_k=A(3-k)\alpha^{2-k}+B\alpha^{3-k}$, where $A \neq 0$.
\end{enumerate}
The first category corresponds to the case when the
 characteristic equation  $a + b \lambda+ c \lambda^2 =0$
has two distinct roots  and we call it the \emph{generic} case.
The second category corresponds to the case when it has a double root
(and $c \not = 0$)
and we call it the \emph{double-root} case.
Category 3 is also a  \emph{double-root} case,
and is only needed for a very special case $b=c=0$.
It can be viewed as the reversal of category 2,
and we always omit the formal proof for this
category since it is similar to  category 2.

For the \emph{generic} case, we can apply a holographic reduction
%under
%the basis
using
$T=\left[
	\begin{array}{c c}
		\alpha_1 & \beta_1 \\
		\alpha_2 & \beta_2
	\end{array}
\right]$. Then we have the following reductions (readers may wish to 
take a look at Section~\ref{useful-reductions} in Appendix):
\begin{eqnarray*}
{\rm Holant}([x_0,x_1,x_2,x_3])
&  \equiv_{\rm T} &
	{\rm Holant}( [1,0,1]|[x_0,x_1,x_2,x_3]) \\
	& \equiv_{\rm T} &
		{\rm Holant}( [1,0,1]T^{\otimes 2} | (T^{-1})^{\otimes 3} [x_0,x_1,x_2,x_3] ) \\
	& \equiv_{\rm T} &
		{\rm Holant}([y_0,y_1,y_2] | [1,0,0,1] ),
\end{eqnarray*}
where $[y_0,y_1,y_2]=[1,0,1]T^{\otimes 2}$. (We note that $[x_0,x_1,x_2,x_3]=T^{\otimes 3}[1,0,0,1]$.)

So for the \emph{generic} case,
we only need to give a dichotomy for ${\rm Holant}([y_0,y_1,y_2] | [1,0,0,1] )$, which
has been proved  in \cite{cai}; we quote their theorem here.

\begin{theorem}\label{lemma-cai} {\rm (\cite{cai})}
The problem Holant$([y_0, y_1, y_2]|[1,0,0,1])$ is \#P-hard for all
$y_0, y_1, y_2 \in \mathbb{C}$ except in the following cases, for which
the problem is in P: (1) $y_1^2=y_0 y_2$;
(2) $y_0^{12}=y_1^{12}$ and $y_0 y_2=-y_1^2$ ( $y_1 \neq 0$);
(3) $y_1=0$; and
(4) $y_0=y_2=0$.
\end{theorem}

To get a complete dichotomy for ${\rm Holant}([x_0,x_1,x_2,x_3])$, we
next deal with the \emph{double-root} case.
%We have the following lemma:

\begin{lemma}\label{lemma-dichotomy-double-reoot}
Let $x_k=A k \alpha^{k-1}+B\alpha^k$, where $A \neq 0$ and $k=0,1,2,3$.
${\rm Holant}([x_0,x_1,x_2,x_3])$ is \#P-hard unless
$\alpha^2 = -1$.
% $\alpha=\pm i (=\sqrt{-1})$,
On the other hand, if $\alpha=\pm i$, then the problem is in P.
\end{lemma}

\begin{proof}
If $\alpha=\pm i$, the signature $[x_0,x_1,x_2,x_3]$ satisfies
 the recurrence relation
$x_{k+2} = \pm 2 i x_{k+1} + x_{k}$, where $k=0,1$. This is a  generalized Fibonacci signature (see \cite{FOCS08}).  Thus we know that it is in P by
holographic algorithms~\cite{FOCS08} using Fibonacci gates.

Now we assume that $\alpha \neq \pm i$. 
%Similar to the proof of Lemma $3.1$ in \cite{STOC09},
We first apply an {\it orthogonal} holographic transformation.
The crucial observation  is that we can view ${\rm Holant}([x_0,x_1,x_2,x_3])$
as the bipartite
${\rm Holant}( [1,0,1]|[x_0,x_1,x_2,x_3])$ and an orthogonal
transformation $T \in {\bf O}_2(\mathbb{C})$ keeps
 $(=_2) = [1,0,1]$
 invariant: $[1,0,1] T^{\otimes 2} = [1,0,1]$.
By a suitable orthogonal transformation $T$, we can  transform
$[x_0,x_1,x_2,x_3]$ to $[v,1,0,0]$ for some $v \in \mathbb{C}$,
up to a scalar. 
(Details are in Appendix.)
So the complexity of ${\rm Holant}([x_0,x_1,x_2,x_3])$ is the same as ${\rm Holant}([v,1,0,0])$.

\begin{figure}[httb]
	\centering
	\includegraphics[height=2.6 cm]{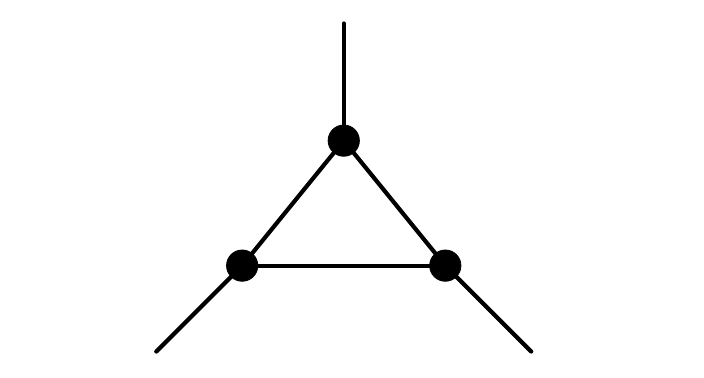}
	\caption{All vertex signatures are $[v,1,0,0]$.}
	\label{fig-g0}
\end{figure}

Next we  prove that ${\rm Holant}([v,1,0,0])$ is \#P-hard for all $v \in \mathbb{C}$.
 %We know that
First, for $v=0$,
Holant($[0,1,0,0]$) is \#P-hard,
because it is the problem of counting all perfect matchings
on 3-regular graphs~\cite{XiaZZ07}.  Second, let  $v \neq 0$.
We can realize $[v^3+3v,v^2+1,v,1]$ by connecting
$[v,1,0,0]$'s as illustrated in Figure \ref{fig-g0},
%%%
%%% JYC:  maybe move the figures around so that this Fig 1
%%%
so it is enough to prove that Holant($[v^3+3v,v^2+1,v,1]$) is \#P-hard.
 In tensor product notation this signature is
\begin{displaymath}
[v^3+3v,v^2+1,v,1]^{\tt T} = \frac{1}{2}\left(
	\left[
		\begin{array}{c}
			v+1 \\
			1
		\end{array}
	\right]^{\otimes 3}
	+\left[
		\begin{array}{c}
			v-1 \\
			1
		\end{array}
	\right]^{\otimes 3}
\right).
\end{displaymath}
Then the following reduction chain holds:
\begin{eqnarray*}
  {\rm Holant}([v^3+3v,v^2+1,v,1])
& \equiv_{\rm T} & {\rm Holant}([1,0,1]|[v^3+3v,v^2+1,v,1]) \\
& \equiv_{\rm T} & {\rm Holant}([v^2+2v+2,v^2,v^2-2v+2]|[1,0,0,1])
\end{eqnarray*}
where the second step is a holographic reduction using
$\left[
        \begin{array}{c c}
                v+1 & v-1 \\
                1 & 1
        \end{array}
\right]$.
We can apply Theorem \ref{lemma-cai} to ${\rm Holant}([v^2+2v+2,v^2,v^2-2v+2]|[1,0,0,1])$, by checking against the four exceptions.
(1) $[v^2+2v+2,v^2,v^2-2v+2]$ is non-degenerate.
(2) There is no solution for
$(v^2-2v+2)^{12}=v^{24}$ and $(v^2+2v+2)(v^2-2v+2)+v^4=0$.
(3) $v^2 \neq 0$. (4) It cannot be the case
%%% JYC:
%%% this point $v^2 \neq 0$ got me confused a little.
%%% had it been v=0, (if we did not seperate out the case v=0)
%%% this would have shown by \ref{lemma-cai} that it is tractable!
%%% this is not a contradiction:  it only says that in 3-regular graphs
%%% if we replace every deg 3 vertex by this gadget (triangle)
%%% then the counting perfect matching problem is indeed in P.
%%% this is using [0 1 0 0] on each vertex of the little triangle.
%%% after holographic transformation. the new deg 3 are [ 1 0 0 1]
%%% and on the edges it's sig [v^2+2v+2,v^2,v^2-2v+2]
%%% which is [2,0,2].
%%%
that $v^2+2v+2=v^2-2v+2=0$.
Therefore
Holant($[v^3+3v,v^2+1,v,1]$) is \#P-hard,
and so is Holant($[v,1,0,0]$) for all $v \in \mathbb{C}$.
\end{proof}

By Theorem \ref{lemma-cai} and Lemma \ref{lemma-dichotomy-double-reoot}, we have
a complete dichotomy theorem for ${\rm Holant}([x_0,x_1,x_2,x_3])$. From
 this, we can get a further
dichotomy for all bipartite Holant problems ${\rm Holant}([y_0,y_1, y_2]|[x_0,x_1,x_2,x_3])$.
The reduction is standard. For any non-degenerate $[y_0, y_1, y_2]$, we can find a transformation $T$,
such that    $[y_0, y_1, y_2]T^{\otimes 2}= [1,0,1]$. Then
the bipartite problem ${\rm Holant}([y_0,y_1, y_2]|[x_0,x_1,x_2,x_3])$
 is  transformed to the equivalent
problem ${\rm Holant}((T^{-1})^{\otimes 3}[x_0,x_1,x_2,x_3])$, for which we have a dichotomy theorem.
%We state the following two theorems
% without spelling out explicitly
%the tractable cases to avoid redundancy. However we note
%that they are both effective and explicit dichotomy theorems.
The following dichotomy theorems are both effective
dichotomies. They use the function
families $\mathscr{A}$ and $\mathscr{P}$,
called {\it affine} functions and functions of a {\it product type}.
(See Section~\ref{Known-Dichotomy} for more details.)

\begin{theorem}\label{thm-ternary}
${\rm Holant}([x_0,x_1,x_2,x_3])$ is \#P-hard unless $[x_0,x_1,x_2,x_3]$ 
satisfies one of the following conditions,
in which case the problem is in P:
\begin{enumerate}
  \item $[x_0,x_1,x_2,x_3]$ is degenerate;
  \item There is a $2\times 2$ matrix $T$ such that $[x_0,x_1,x_2,x_3]=T^{\otimes 3} [1,0,0,1]$ and $[1,0,1]T^{\otimes 2}$ is in $\mathscr{A} \cup \mathscr{P}$;
  \item For $\alpha \in \{2i, -2i\}$,
 $x_2+\alpha x_1 -x_0=0$ and $x_3+\alpha x_2 -x_1=0$.
\end{enumerate}
\end{theorem}

\begin{theorem}
${\rm Holant}([y_0,y_1, y_2]|[x_0,x_1,x_2,x_3])$ is \#P-hard unless $[x_0,x_1,x_2,x_3]$ and $[y_0,y_1, y_2]$ satisfy one of the following conditions,
in which case the problem is in P:
\begin{enumerate}
  \item $[x_0,x_1,x_2,x_3]$ is degenerate;
  \item There is a $2\times 2$ matrix $T$ such that $[x_0,x_1,x_2,x_3]=T^{\otimes 3} [1,0,0,1]$ and $[y_0,y_1, y_2]T^{\otimes 2}$ is in $\mathscr{A} \cup \mathscr{P}$;
  \item There is a $2\times 2$ matrix $T$ such that $[x_0,x_1,x_2,x_3]=T^{\otimes 3} [1,1,0,0]$ and $[y_0,y_1, y_2]T^{\otimes 2}$ is of form $[0,*,*]$;
  \item There is a $2\times 2$ matrix $T$ such that $[x_0,x_1,x_2,x_3]=T^{\otimes 3} [0,0,1,1]$ and $[y_0,y_1, y_2]T^{\otimes 2}$ is of form $[*,*,0]$.
\end{enumerate}
\end{theorem}

%% file: holant-to-CSP.tex
\section{Reductions Between Holant and \#CSP}\label{sec-holant-to-csp}
%%% JYC in this sec, i also changed F_1 to G_1 , F_2 to G_2
%%% to avoid collision with the fixed F1 F2 F3 family.
%%%
In this section, we extend the dichotomies in Section~\ref{sec-ternary}
 for a single ternary signature to a set of signatures. We are
%%%%% JYC:  for notational simplicity I suggeste we use :
%%%     ${\rm Holant}( [x_0,x_1,x_2,x_3] \cup \mathscr{F})$
%%% for ${\rm Holant}(\{ [x_0,x_1,x_2,x_3] \} \cup \mathscr{F})$
going to give a dichotomy 
for ${\rm Holant}([x_0,x_1,x_2,x_3]  \cup \mathscr{F})$, or more generally for
${\rm Holant}( [y_0,y_1, y_2]  \cup \mathscr{G}_1 | [x_0,x_1,x_2,x_3] \cup \mathscr{G}_2)$,
where $[y_0,y_1, y_2]$ and $[x_0,x_1,x_2,x_3]$ are non-degenerate.
In this section, we focus on the generic case of $[x_0,x_1,x_2,x_3]$, 
and the double root case will be
handled in the next section in Lemma \ref{lemma-double-root}. 
%%% JYC: I think I know what you mean here. all of this is geared toward
%%% the Holant^c.  but for the "Reductions Between Holant and \#CSP"
%%% as is, this Lemma 5.2 in sec 5 ( Lemma \ref{lemma-double-root)
%%% is not exactly matched up here.
For the generic case, we can apply a holographic reduction
to transform  $[x_0,x_1,x_2,x_3]$ to $[1,0,0,1]$. 
%Therefore we only need to give a dichotomy for the normalized problems
%%% JYC:  at this point calling it "normalized" will be confusing.
Therefore we only need to give a dichotomy for Holant problems of the form
Holant$( [y_0,y_1, y_2]  \cup \mathscr{G}_1 | [1,0,0,1] \cup \mathscr{G}_2 )$, where $[y_0,y_1, y_2]$ is non-degenerate.
We make one more observation: The ternary equality signature 
$[1,0,0,1]$ is invariant under the following transformations:
\begin{displaymath}
\mathscr{T}_3 \triangleq \left\{
	\left[
		\begin{array}{c c}
			1 & 0 \\
			0 & 1
		\end{array}
	\right],
	\left[
		\begin{array}{c c}
			1 & 0 \\
			0 & \omega
		\end{array}
	\right],
	\left[
		\begin{array}{c c}
			1 & 0 \\
			0 & \omega^2
		\end{array}
	\right]
	\right\},
\end{displaymath}
where $\omega= \omega_3 = e^{2 \pi i/3}$. 
%So we can apply a holographic reduction under a 
%basis $T \in \mathscr{T}_3 $ to further normalize the problem if we find
%it convenient.
For any $T \in \mathscr{T}_3 $,
\[ {\rm Holant}([y_0, y_1, y_2]  |  [1,0,0,1] \cup \mathscr{F} ) 
\equiv_{\rm T} {\rm Holant}([y_0, y_1, y_2]T^{\otimes 2}  | [1,0,0,1] \cup T^{-1}\mathscr{F} ) .\]
As a result, we can normalize $[y_0, y_1, y_2]$ by 
a holographic reduction with any $T\in \mathscr{T}_3 $. In particular, we call a symmetric
binary signature $[y_0, y_1, y_2]$ \emph{normalized} if $y_0=0$ or it is 
not the case that
% $\frac{y_2}{y_0}$ ($y_0 \neq 0$)
$y_2$ is $y_0$ times a $t$-th primitive root of unity,
 and $t=3t'$ where ${\rm gcd}(t', 3) =1$. 
If $[y_0, y_1, y_2]$ is not normalized, then $y_2=y_0 \omega_{t}^s$,
where $\omega_{t} = e^{2 \pi i/t}$ and
 ${\rm gcd}(s, t) =1$.  Write $1 = 3 u + t' v$ for some integers
$u$ and $v$, then $\omega_{t} = \omega^v \omega_{t'}^u$
and $\omega_{t}^s = \omega^k \omega_{t'}^l$,
where $k \equiv sv \bmod 3$, and ${\rm gcd}(k, 3) =1$.
Hence $k =1$ or $2$.
After applying the transformation  	
$\left[		\begin{array}{c c}
			1 & 0 \\
			0 & \omega^k
		\end{array}
	\right] \in \mathscr{T}_3 $,
we get a new signature $[y_0, y_1 \omega^k , y_0 \omega_{t'}^l]$, which is normalized. So in the following, we only deal with normalized
$[y_0, y_1, y_2]$. In one case, we also need to normalized a unary signature
$[x_0, x_1]$, namely $x_0=0$ or $x_1$ is not a multiple of $x_0$
by a $t$-th primitive root of unity, and $t=3t'$ where ${\rm gcd}(t', 3) =1$.
%  $[1,a]$, where $a \neq 0$.
Again we can normalize the unary signature by a suitable $T\in \mathscr{T}_3$.
%Similar, we can normalize the unary signature by a 
%transform  $T\in \mathscr{T}_3 $ such that it is not the 
%case that $a$ is a $t$-th primitive root of unity and $t=3t'$, 
%where $3 \not | t'$.  We call such unary signature $[1,a]$ normalized.
We note that a normalized signature is still normalized after a scalar
multiple.

\begin{theorem}\label{thm-bipartite}
Let $[y_0, y_1, y_2]$ be a normalized and non-degenerate signature. And in the case of $y_0=y_2=0$, we further assume
that $\mathscr{G}_1$ contains a unary signature $[a,b]$, which is normalized and  $a b \neq 0$.
Then
\[ {\rm Holant}(  [y_0,y_1, y_2]  \cup \mathscr{G}_1 
|  [1,0,0,1] \cup \mathscr{G}_2) \equiv_{\rm T} \#{\rm CSP}([y_0, y_1, y_2]\cup \mathscr{G}_1 \cup \mathscr{G}_2).\]
More specifically,  ${\rm Holant}( [y_0,y_1, y_2] \cup \mathscr{G}_1
 |  [1,0,0,1] \cup \mathscr{G}_2)$ is \#P-hard unless
$[y_0, y_1, y_2]\cup \mathscr{G}_1 \cup \mathscr{G}_2 \subseteq \mathscr{P}$ or $[y_0, y_1, y_2]\cup \mathscr{G}_1 \cup \mathscr{G}_2 \subseteq \mathscr{A}$,
in which cases the problem is in P.
\end{theorem}

This dichotomy is an important reduction step in the proof of our
dichotomy theorem for Holant$^c$. It is also interesting in its own right
as a connection between Holant and \#CSP.
The assumption on signature normalization
in the statement of the theorem is without loss of generality. 
For a non-normalized signature, we can apply a
normalization and then apply the dichotomy criterion in the theorem. 
The additional assumption of the existence of a non-zero unary signature 
circumvents a technical difficulty, and finds a circuitous
route
to the proof of our dichotomy theorem for Holant$^c$, the main
objective in this paper.  For Holant$^c$,
the needed unary signature will be produced  from $[1,0]$ and $[0,1]$.
%The additional assumption of the existence of a non-zero unary signature is 
%technical, but will be  
%This version is sufficient to get a dichotomy theorem for Holant$^c$, 
%where the unary will be produced  from $[1,0]$ and $[0,1]$.
%%% JYC you had "$[1,0]$ or $[0,1]$" but i think you may need both
%%% to get it, when you take sub-signatures.
%However, it would be nice if we can remove this assumption and thus get a 
%more generic dichotomy. We leave this as an interesting open question.
We also note that we do not require the signatures in  $\mathscr{G}_1$ and  $\mathscr{G}_2$ to be symmetric, so the theorem could have applications
in dichotomy theorems for general Holant problems over non-symmetric signatures.

One direction in Theorem~\ref{thm-bipartite}, from Holant to \#CSP, is
straightforward.
Thus our main claim is a reduction from  \#CSP to these bipartite
Holant problems.
The approach is to construct the binary equality gate $[1,0,1] = (=_2)$
in LHS in the Holant problem.
As soon as we have $[1,0,1]$ in LHS, together with $[1,0,0,1]
= (=_3)$ in RHS, we get equality gates of all
arities $(=_k)$ in RHS. Also with the help of $[1,0,1]$ in LHS
we can transfer $\mathscr{G}_2$ to LHS. Then we have all of
$\#{\rm CSP}([y_0, y_1, y_2]\cup \mathscr{G}_1 \cup \mathscr{G}_2)$.

If the problem Holant$([y_0, y_1, y_2]  | [1,0,0,1]  )$ is already \#P-hard,
then for any $\mathscr{G}_1$ and  $\mathscr{G}_2$, it is \#P-hard. So we only need to consider the cases, where
Holant$([y_0, y_1, y_2]  | [1,0,0,1]  )$ is not \#P-hard.
%%% JYC: technically we don't have dichotomy yet...of course we can by thm 3.1
%%% is polynomial computable.
For this, we again use Theorem \ref{lemma-cai} from \cite{cai}.
The first tractable case $y_1^2=y_0 y_2$ is degenerate, 
which does not apply here.
%which is not interesting here. 
The following three lemmas deal with the remaining three tractable cases respectively.

For the case $y_0^{12}=y_1^{12}$ and $y_0 y_2=-y_1^2$ ( $y_1 \neq 0$), we can scale it to $[a,1,b]$, where
$a^{12}=1$ and $ab=-1$. Since $[a,1,b]$ is normalized,
%it can be verified that we have further   $a^4=1$.
it follows that $a^4=1$.
 %Then we have the following lemma.

\begin{lemma}
Let $\mathscr{G}_1$ and  $\mathscr{G}_2$  be two sets of signatures.
For all pairs of $a$ and $b$ satisfying $a^{4}=1$ and
$ab=-1$, ${\rm Holant}([a,1,b]\cup\mathscr{G}_1 |[1,0,0,1]\cup\mathscr{G}_2)$ is \#P-hard unless
$\mathscr{G}_1 \cup \mathscr{G}_2 \subseteq \mathscr{A} $, in which case it is in P.
\end{lemma}

%
%\begin{lemma}\label{lemma-equal-a1b}
%Let $\mathscr{F}$ be a set of signatures.
%For all pairs of $a$, $b$ satisfying $a^{4}=1$ and
%$ab=-1$, we have
%\begin{eqnarray*}
%{\rm Holant}(\{[a,1,b]\}\cup\mathscr{F}|\{=_3\}) & \equiv_{\rm T} &
%	{\rm Holant}(\{[a,1,b],=_2\}\cup\mathscr{F}|\{=_3,=_2\}) \\ & \equiv_{\rm T} &
%	{\rm Holant}(\{[a,1,b],=_3\}\cup\mathscr{F}) \\ & \equiv_{\rm T} &
%	\rm{\#CSP}(\{[a,1,b]\}\cup\mathscr{F})
%\end{eqnarray*}
%\end{lemma}
\begin{proof}
We first prove that when $a^{4}=1$ and
$ab=-1$,
\[{\rm Holant}( [a,1,b]\cup\mathscr{G}_1 |[1,0,0,1]\cup\mathscr{G}_2) 
\equiv_{\rm T} \#{\rm CSP}([a,1,b]\cup\mathscr{G}_1 \cup \mathscr{G}_2).\]
To get this, it is sufficient to construct $[1,0,1]$ in LHS.

\begin{figure}[httb]
	\begin{minipage}[t]{0.48\linewidth}
		\centering
		\includegraphics[height=3.7 cm]{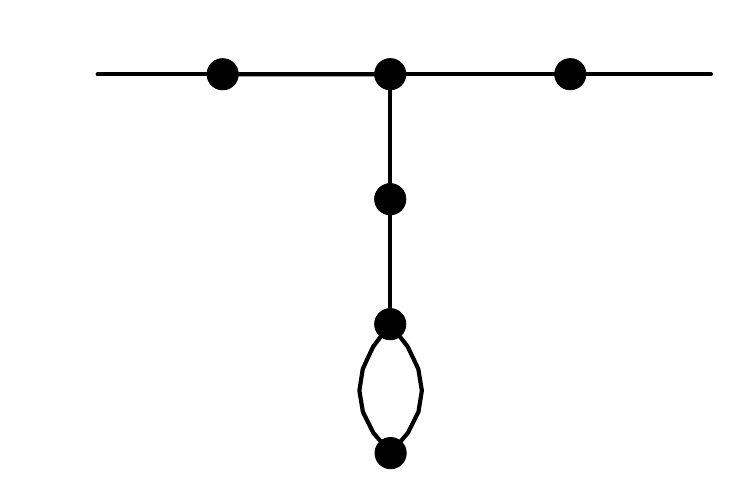}
		\caption{A gadget construction
for the
binary disequality gate $(\not=_2)$ on LHS.
			Degree 3 vertices have
 signature  $=_3$, degree 2 vertices  have
  signature $[1,\pm i,1]$.}
		\label{fig-gt1}
	\end{minipage}
	\hspace{0.02\linewidth}
	\begin{minipage}[t]{0.48\linewidth}
		\centering
		\includegraphics[height=3.7 cm]{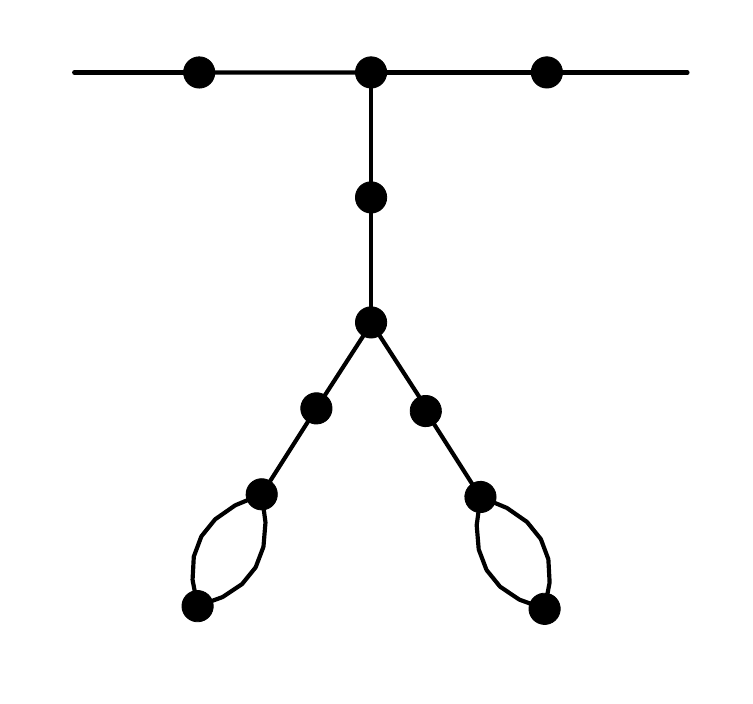}
		\caption{Another gadget construction
for $(\not=_2)$ on LHS. 
Degree 3 vertices have
 signature  $=_3$, degree 2 vertices 
  %signature $[1,\pm i,1]$.}
signature $[1,\pm1,-1]$.}
		\label{fig-gt2}
	\end{minipage}
\end{figure}

\noindent 
\emph{Case 1}: $a=\pm i$.

It is equivalent to
consider ${\rm Holant}([1,\pm i,1] \cup\mathscr{G}_1 |[1,0,0,1]\cup\mathscr{G}_2)$ because
they only differ by a constant factor.
We can construct $[1,1]$ on RHS by connecting the two edges of a
 $[1,\pm i,1]$ gate on the LHS
with two edges of a $[1,0,0,1] = (=_3)$ on the RHS. With the gadget
in Figure \ref{fig-gt1}, we can construct the binary disequality gate 
 $[0,1,0] = (\not =_2)$ on LHS. Together with
%%% JYC: Again pl rearrange the order of the Figures.
the $=_3$ on RHS, we can have $=_3$ on LHS.
%%% because the on ly way to \not = in {0,1} is to equal the other value.
Connecting this LHS $=_3$ with a  $[1,1]$ on RHS, we can obtain 
the binary equality gate $[1,0,1] = (=_2)$ on LHS.

\noindent
\emph{Case 2}: $a=\pm 1$.

It is equivalent to
consider ${\rm Holant}([1,\pm 1,-1]\cup\mathscr{G}_1 
|[1,0,0,1]\cup\mathscr{G}_2)$.
With the gadget in Figure \ref{fig-gt2}, we can construct $[0,1,0]$ on LHS, and
thus $[1,0,0,1]$ on LHS. Furthermore, we can construct
$[1,-1]$ and $[1,0,-1]$ on both sides. By connecting $[1,-1]$ with $[1,0,-1]$,
we can realize $[1,1]$ on both sides, and consequently $[1,0,1]$
on both sides.
%%%% JYC: I have yet to check that.

\vspace{.1in}
By~\cite{STOC09}, 
 $\#{\rm CSP}([a,1,b]\cup\mathscr{F})$ is \#P-hard
unless $[a,1,b]\cup \mathscr{G}_1 \cup \mathscr{G}_2 \subseteq \mathscr{P} $ 
or $[a,1,b]\cup\mathscr{G}_1 \cup \mathscr{G}_2 \subseteq \mathscr{A} $.
Since $[a,1,b] \in \mathscr{A} - \mathscr{P}$, we conclude that
 the only possible case which is not \#P-hard is
$\mathscr{G}_1 \cup \mathscr{G}_2 \subseteq \mathscr{A} $. 
This is also sufficient  for tractability.
The proof is complete.
%This completes the proof.
\end{proof}

For the tractable case $y_1=0$ in Theorem~\ref{lemma-cai}, 
by non-degeneracy,
we can scale it to be $[1,0,a]$, where $a \neq 0$.
Then  we have the following lemma:

\begin{lemma}\label{lemma-equal-10a}
Let $\mathscr{G}_1$ and  $\mathscr{G}_2$  be two sets of signatures, 
and let $a \neq 0$ be a complex
number. We assume $[1,0,a]$ is normalized. Then we have the following dichotomy:
\begin{itemize}
  \item If $a^4=1$, then ${\rm Holant}([1,0,a] \cup\mathscr{G}_1 
|[1,0,0,1]\cup\mathscr{G}_2)$ is \#P-hard unless
$\mathscr{G}_1 \cup \mathscr{G}_2 \subseteq \mathscr{P} $ or $\mathscr{G}_1 \cup \mathscr{G}_2 \subseteq \mathscr{A} $, in which cases it is in P.
  \item if $a^4 \neq 1$, then ${\rm Holant}([1,0,a] \cup\mathscr{G}_1 
|[1,0,0,1]\cup\mathscr{G}_2)$ is \#P-hard unless
$\mathscr{G}_1 \cup \mathscr{G}_2 \subseteq \mathscr{P} $,  in which case it is in P.
\end{itemize}
\end{lemma}
\begin{figure}[httb]
	\begin{minipage}[t]{.48\textwidth}
		\includegraphics[width=.9\textwidth]{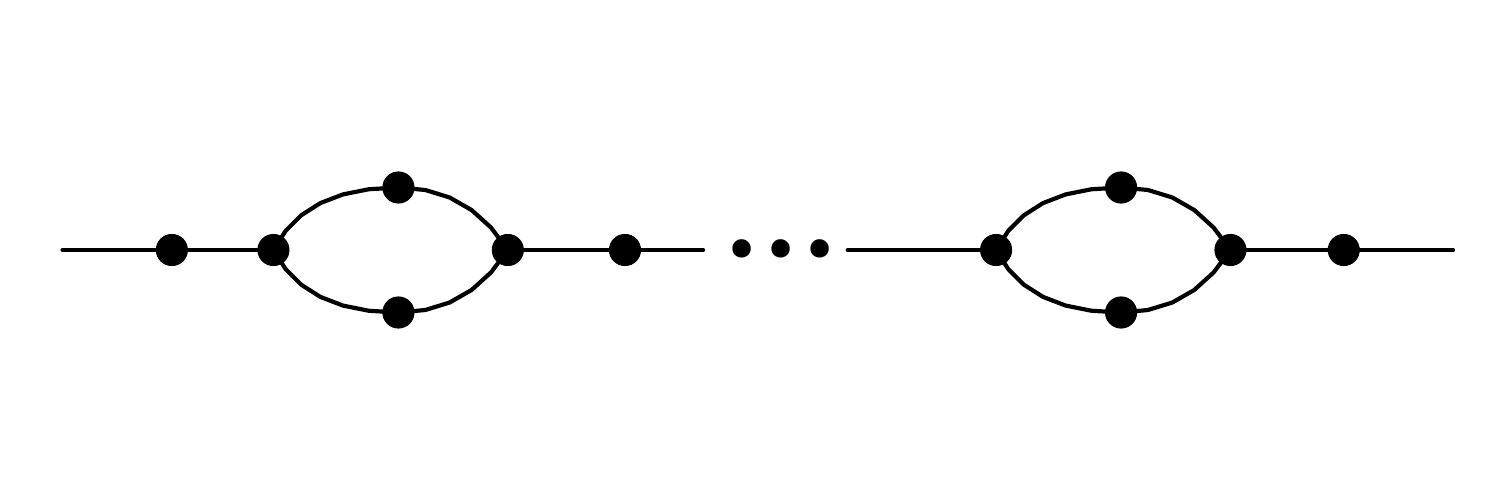}
		\caption{A recursive gadget we use to construct $[1,0,a^{3k+1}]$ on LHS.
			Ternary signatures are $[1,0,0,1]$, and binary signatures are
			$[1,0,a]$.}
		\label{fig-gp1}
	\end{minipage}
	\hspace{.02\linewidth}
	\begin{minipage}[t]{.48\textwidth}
		\centering
		\includegraphics[width=.9\textwidth]{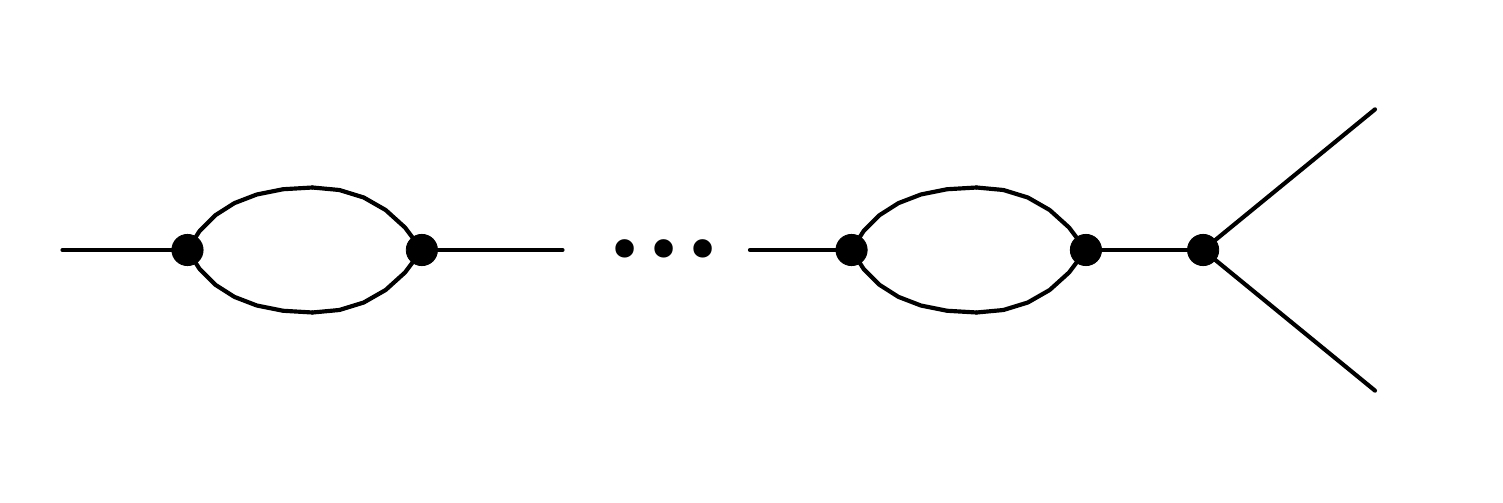}
\caption{A recursive gadget we use to realize 
%$[1,0,0,-\omega_{3^{(l-1)}}^{mbk_0/2}]$
 $[1,0,0, \omega_{3^{(l-1)}}^{mbk_0}]$
			where $m$ is an odd integer. All ternary gadgets are
%			$[1,0,0,-\omega_{3^{(l-1)}}^{bk_0/2}]$.}
                       $[1,0,0,  \omega_{3^{(l-1)}}^{bk_0}]$.}
		\label{fig-gp2}
	\end{minipage}
\end{figure}
%%% JYC:  I think you still have to include [1,0,a] together with F_1 and F_2
%%% to be subset of P or A...

%\begin{enumerate}
%\item If $a$ is not a root of unity, then
%	\begin{displaymath}
%		{\rm Holant}(\{[1,0,a]\}|\{[1,0,0,1]^T\}\cup\mathscr{F})
%			\equiv_{\rm T} \rm{\#CSP}(\{[1,0,a]\}\cup\mathscr{F})
%	\end{displaymath}
%\item If $a$ is a $t$-th root of unity, and $t$ is not a multiple of 3, then
%	\begin{displaymath}
%		{\rm Holant}(\{[1,0,a]\}|\{[1,0,0,1]^T\}\cup\mathscr{F})
%			\equiv_{\rm T} \rm{\#CSP}(\{[1,0,a]\}\cup\mathscr{F})
%	\end{displaymath}
%\item If $a$ is a $t$-th root of unity, and the power of $3$ in $t$ is
%	greater than 1, then
%	\begin{eqnarray*}
%		& & {\rm Holant}(\{[1,0,a]\}|\{[1,0,0,1]^T\}\cup\mathscr{F}) \\
%		& \equiv_{\rm T} & \rm{\#CSP}([1,0,a](T^{-1})^{\otimes 2},
%				[1,0,0,1]T^{\otimes 3},T\mathscr{F})
%	\end{eqnarray*}
%	for some suitable diagonal matrix $T$;
%\item Otherwise, we have
%	\begin{displaymath}
%		{\rm Holant}([1,0,a]|[1,0,0,1],\mathscr{F})
%			\equiv_{\rm T} \rm{\#CSP}([1,0,a](T^{-1})^{\otimes 2},T\mathscr{F})
%	\end{displaymath}
%	for some suitable diagonal matrix $T$.
%\end{enumerate}
%\end{lemma}
\begin{proof}
As above, 
%to prove \#P-hardness
it is sufficient to show that we can construct $[1,0,1]$ in LHS.
We will use the gadget in Figure \ref{fig-gp1} in our proof. We can use it to
realize $[1,0,a^{3k+1}]$ for any $k \in \mathbb{N}$ on LHS.

If $a$ is not a root of unity, then we will be able to interpolate all
signatures of the form $[1,0,x]$ where $x \in \mathbb{C}$ on LHS. 
This uses a Vandermonde system and we omit the details. In particular,
we will be able to interpolate $[1,0,1]$ on LHS. So we are done.
%%% JYC: technically we should say, since a is not a root of unity
%%% bullet 1 in thm 4.3 does not apply. bullet 2 has a problem as
%%% stated above.
%%% also must say the affine case does not apply in this case of "$a$ is 
%%% not a root of unity".

Now we can assume that $a$ is a $t$-th primitive root of unity, 
that is $a=\omega_{t}^b$ for some
$b$ relatively prime to $t$, where $\omega_{t}=e^{2\pi i/t}$. 
If $t$ is not a multiple
of 3, then we can find an integer $k$, such that $3k+1 \equiv 0 \pmod{t}$.
Therefore, we can realize $[1,0,1]$ on LHS, and carry out the same reduction
as above.
%%% JYC: exactly what is above: the part above here in lm 4.3 or in lm 4.2?

Now we consider the case of $t=3^l t'$, where
$l \ge 1$ and ${\rm gcd}(t', 3) =1$. 
%$t'$ is not a multiple
%of 3. 
Since $[1,0,a]$ is normalized, we have a further condition that $l>1$.
For this case, we do not know how to construct $[1,0,1]$ in LHS directly.
Instead we will
further apply a holographic reduction. Also in this case, we have $a^4 \neq 1$, so we want to
prove that ${\rm Holant}([1,0,a] \cup\mathscr{G}_1 
|[1,0,0,1]\cup\mathscr{G}_2)$ is \#P-hard unless
$\mathscr{G}_1 \cup \mathscr{G}_2 \subseteq \mathscr{P} $. The fact that the problem is in P when $\mathscr{G}_1 \cup \mathscr{G}_2 \subseteq \mathscr{P} $
is obvious by Theorem~\ref{thm:dichotomy}, since $[1,0,a] \in \mathscr{P}$.

Since $2t'$ is  not a multiple
of 3, there exist some integers $k$ and $k_0$,  such that
%\begin{displaymath}
$3k+1=2 k_0t'$.
%\end{displaymath}
Since
\begin{displaymath}
a^{3k+1}=a^{2 k_0t'}=\omega_{3^lt'}^{2 b k_0t'}=\omega_{3^l}^{2bk_0}
\end{displaymath}
we can realize $[1,0,\omega_{3^l}^{2 bk_0}]$ on LHS.
So
\[ {\rm Holant}([1,0,\omega_{3^l}^{2 bk_0}] \cup\mathscr{G}_1 
|\{=_3\}\cup\mathscr{G}_2) \leq_T {\rm Holant}([1,0,a] \cup\mathscr{G}_1 
|[1,0,0,1]\cup\mathscr{G}_2). \]
Therefore it is sufficient to prove that 
${\rm Holant}([1,0,\omega_{3^l}^{2 bk_0}] \cup\mathscr{G}_1 
|\{=_3\}\cup\mathscr{G}_2)$ is \#P-hard
if $\mathscr{G}_1 \cup \mathscr{G}_2 \not \subseteq \mathscr{P} $.

We apply a holographic reduction under the basis $
T=\left[
\begin{array}{c c}
	1 & 0 \\
	0 & \omega_{3^l}^{-bk_0}
\end{array}
\right]
$, and get 
$${\rm Holant}([1,0,\omega_{3^l}^{2 bk_0}] \cup\mathscr{G}_1 
|\{=_3\}\cup\mathscr{G}_2) \equiv_{\rm T} {\rm Holant}( [1,0,1] \cup\mathscr{G}_1 T 
|[1,0,0,\omega_{3^{(l-1)}}^{bk_0}]\cup T^{-1} \mathscr{G}_2).$$

We then use the gadget in Figure \ref{fig-gp2} to realize
 $[1,0,0,\omega_{3^{(l-1)}}^{3^{l-1}bk_0}]=[1,0,0,1] = (=_3)$ in RHS.
Together with $[1,0,1] = (=_2)$ in LHS this gives all
equality gates.
As a result
$$\#{\rm CSP}([1,0,0,\omega_{3^{(l-1)}}^{bk_0}] \cup\mathscr{G}_1 T 
\cup T^{-1} \mathscr{G}_2)
 \leq_T {\rm Holant}( [1,0,1] \cup\mathscr{G}_1 T
 |[1,0,0,\omega_{3^{(l-1)}}^{bk_0}]\cup T^{-1} \mathscr{G}_2).$$
Since $l>1$, $[1,0,0,\omega_{3^{(l-1)}}^{bk_0}] \not \in \mathscr{A}$.
So the problem is \#P-hard unless  $\mathscr{G}_1 T \cup T^{-1} \mathscr{G}_2
 \subseteq \mathscr{P}$.
Since $T$ and $T^{-1}$ are diagonal matrices, 
it is equivalent to  say that $ \mathscr{G}_1 \cup \mathscr{G}_2 \subseteq \mathscr{P}$.
This completes the proof.

\end{proof}

For the last tractable case $y_0=y_2=0$
 in Theorem~\ref{lemma-cai}, we can scale it to $[0,1,0]$. 
%Here we prove a 
%slightly weaker dichotomy, which assumes that we also have a unary 
%signature $[1,a]$ in $\mathscr{G}_1$ (or $\mathscr{G}_2$), where $a \neq 0$.

\begin{lemma}\label{lemma-equal-010}
Let $\mathscr{G}_1$ and  $\mathscr{G}_2$  be two sets of signatures, and $a \neq 0$ be a complex
number. We assume $[1,a]$ is normalized.  Then we have the following dichotomy:
\begin{itemize}
  \item If $a^4=1$, then ${\rm Holant}(\{[0,1,0],[1,a]\} \cup  \mathscr{G}_1 |[1,0,0,1] \cup \mathscr{G}_2)$ is \#P-hard unless
$\mathscr{G}_1 \cup \mathscr{G}_2 \subseteq \mathscr{P} $ or $\mathscr{G}_1 \cup \mathscr{G}_2 \subseteq \mathscr{A} $, in which cases it is in P.
  \item if $a^4 \neq 1$, then ${\rm Holant}(\{[0,1,0],[1,a]\} \cup  \mathscr{G}_1 |[1,0,0,1] \cup \mathscr{G}_2)$ is \#P-hard unless
$\mathscr{G}_1 \cup \mathscr{G}_2 \subseteq \mathscr{P} $,  in which case it is in P.
\end{itemize}
\end{lemma}
\begin{proof}
By connecting a $[a,1]$ and two $[0,1,0]$'s to a $[1,0,0,1]$, we can realize $[a,0,1]$, or
equivalently $[1,0,1/a]$ on LHS, and by Lemma \ref{lemma-equal-10a}, the proof follows.
\end{proof}

%To sum up these three lemmas, we get the following dichotomy theorem for Holant$([y_0, y_1, y_2]  | \{ [1,0,0,1] \} \cup \mathscr{F} )$.
%
%\begin{theorem}\label{thm-bipartite}
%Let $[y_0, y_1, y_2]$ be normalized and non-degenerate. And in the case of $y_0=y_2=0$, we further assume
%that $\mathscr{F}$ contains a unary signature $[a,b]$, which is normalized and  $a b \neq 0$.
%Then Holant$([y_0, y_1, y_2]  | \{ [1,0,0,1] \} \cup \mathscr{F} ) \equiv_{\rm T} \rm{\#CSP}([y_0, y_1, y_2]\cup \mathscr{F}$.
%Or more specified,  Holant$([y_0, y_1, y_2]  | \{ [1,0,0,1] \} \cup \mathscr{F} )$ is \#P-hard unless
%$[y_0, y_1, y_2]\cup \mathscr{F} \subseteq \mathscr{P}$ or $[y_0, y_1, y_2]\cup \mathscr{F} \subseteq \mathscr{A}$,
%in which cases the problem is in P.
%\end{theorem}

%% file: dichotomy.tex
\section{Dichotomy Theorem for Complex Holant$^c$ Problems}
In this section, we prove our main result, a dichotomy theorem for Holant$^c$ problems with complex valued symmetric signatures, 
which is stated as Theorem \ref{thm:main}. The proof crucially uses
 the dichotomies proved in the previous two sections. 
In order to use them, we first prove in Lemma \ref{lm-exist-ternary} that we can always realize a non-degenerate ternary 
signature except
in some trivial cases. After having a non-degenerate ternary
signature, we can immediately prove \#P-hard if the ternary
is not of one of the tractable cases in Theorem \ref{thm-ternary}. For the tractable ternary signatures, we use Theorem \ref{thm-bipartite}
to extend the dichotomy theorem to the whole signature set. In Theorem \ref{thm-bipartite}, we only consider the generic case 
of the ternary function. The double-root case is handled here in Lemma \ref{lemma-double-root}.

\begin{lemma}\label{lm-exist-ternary}
Given any set of symmetric signatures $\mathscr{F}$ which contains $[1,0]$
 and $[0,1]$, we can construct a non-degenerate symmetric ternary signature
$X=[x_0,x_1,x_2,x_3]$, except in the following two trivial cases:
\begin{enumerate}
  \item Any non-degenerate signature in $\mathscr{F}$ is of arity at most $2$;
  \item In $\mathscr{F}$, all unary signatures are of form $[x,0]$ or $[0,x]$;
  all binary signatures are of form $[x,0,y]$ or $[0,x,0]$;
  and all signatures of arity greater
 than 2 are of form $[x,0,\ldots, 0, y]$.
\end{enumerate}
\end{lemma}

\begin{proof}
Suppose case 1. does not hold, and let
$[x_0,x_1, \ldots, x_m] \in \mathscr{F}$ be
a non-degenerate signature of arity at least 3.
Since we have $[1,0]$ and $[0,1]$,
we can construct all sub-signatures of any signature in   $\mathscr{F}$.
%%% JYC: need to define them in background section
If there exists  a ternary non-degenerate sub-signature, we are done.
Now suppose all ternary sub-signatures are degenerate, and $m >3$.
%If all ternary sub-signatures of it are degenerate,
Then we can show that it must be of
form $[x_0, 0, \ldots, 0, x_m]$, where $x_0 x_m \neq 0$.
%%% JYC: this actually need a proof in the final verison to journal
%%% say some x_i \not = 0, for some 0< i< m.
%%% if all x_j, are non-zero, and , can divide, and it forms geometric
%%% series. so either some x_j  =0 for some j<i, or for some j>i.
%%% take the first place. and consider the 2 by 2 matrix. contradict
%%% degeneracy there.
%So we only need to consider the case that all signatures in
%$\mathscr{F}$ of arity at least 3 are of form $[x,0,\ldots, 0, y]$.
% There is at least one signature with this form,
%otherwise we are in the exceptional case 1.
If we have a unary signature $[a,b]$ where $a b \neq 0$ or
a unary sub-signature $[a,b]$ (where $a b \neq 0$) of a binary signature, we can connect this signature to $m-3$ dangling
edges of   $[x_0,  0, \ldots, 0, x_m]$
to get a non-degenerate ternary signature $[x,0,0,y]$, and we are done.
Otherwise, all unary signatures are of form $[x,0]$ or $[0,x]$ and
  all binary signatures are of form $[x,0,y]$ or $[0,x,0]$. Therefore
 we are in the exceptional case 2.
\end{proof}

We next consider the double root case for $X=[x_0,x_1,x_2,x_3]$.
By Lemma~\ref{lemma-dichotomy-double-reoot}, $\textrm{Holant}(X)$
is already \#P-hard unless  the double eigenvalue is 
$i$ or $-i$.
%In this case, we have
Then, 
$x_{k+2}  + \alpha x_{k+1}-x_{k}=0$ for $k=0,1$, where $\alpha = \pm 2i$. 
\begin{lemma}\label{lemma-double-root}
Let $X=[x_0,x_1,x_2,x_3]$ be a complex signature satisfying
$x_{k+2}  + \alpha x_{k+1}-x_{k}=0$ for $k=0,1$, where $\alpha = \pm 2i$.
 Let $Y=[y_0,y_1,y_2]$
be a non-degenerate binary signature. Then $\textrm{Holant}(\{X,Y\})$
%%% {}
is \#P-hard unless $y_2 + \alpha y_1-y_0=0$ (in which case
it is in P by Fibonacci gates).
%%% JYC: my change is to say, the pm matches up in \alpha = \pm 2i (in X and Y)
\end{lemma}
\begin{proof}
We prove this result for $\alpha = -2i$. The other case is similar.
% $x_{k+2}-2ix_{k+1}-x_{k}=0$. The other case is similar.

The sequence $\{x_k\}$ can be written as follows:
$x_k=Aki^{k-1}+Bi^{k}$, where $ A \neq 0$.
Thus, we have
\begin{displaymath}
	X=T^{\otimes 3}[1,1,0,0]^{\rm T},
~~~~\text{ where }
	T=\left[
		\begin{array}{c c}
			1 & \frac{B-1}{3} \\
			i & A+\frac{B-1}{3}i
		\end{array}
	\right].
\end{displaymath}
By expressing 
$\left[
                \begin{array}{c c}
                        y_0 & y_1 \\
                        y_1 & y_2                \end{array}
        \right]
=T_0^{\tt T} T_0$, which is always possible for some non-singular
$	T_0=\left[
		\begin{array}{c c}
			a & c \\
			b & d
		\end{array}
	\right]
$, we have
\[Y  =  [1,0,1]T_0^{\otimes 2} = 
(\left( \begin{array}{c c} 1 & 0 \end{array} \right)T_0)^{\otimes 2} + 
(\left( \begin{array}{c c} 0 & 1 \end{array} \right)T_0)^{\otimes 2}
		 =  [a^2+b^2,ac+bd,c^2+d^2].\]
%%% JYC:
%%% if y_0 not =0, then do 
%%% (1 0; -y_1/y_0  1) (y_0  y_1; y_1  y_2) (1 -y_1/y_0; 0  1) = diagonal
%%% if y_0=0,i but y_2 not =0, then use switch (0 1;1 0)
%%% if y_0=y_2=0. then y_1 \not =0.
%%% then (1 1; 0 1) (0  y_1; y_1 0) (1 0; 1 1) = (2y_1 y_1; y_1 0)
%%% now the new y_0 not =0

Thus we have the following chain of reductions
\begin{comment}\begin{eqnarray}\label{T_0-and-T}\end{comment}
\begin{displaymath}
\textrm{Holant}(Y|X)
 \equiv_{\rm T}  \textrm{Holant}([1,0,1]T_0^{\otimes 2}|T^{\otimes 3}[1,1,0,0]^{\tt T})
 \equiv_{\rm T}  \textrm{Holant}([1,0,1]|(T_0T)^{\otimes 3}[1,1,0,0]^{\tt T}),
%~~~\mbox{   where,}
\end{displaymath}
%\begin{comment}\end{eqnarray}\end{comment}
%We write
\begin{eqnarray*}
\mbox{   where,}~~~
T_0T & = &
\left[\begin{array}{c c}
a+ci & Ac+\frac{B-1}{3}(a+ci) \\
b+di & Ad+\frac{B-1}{3}(b+di)
\end{array}\right], ~~~~\text{ and we will call it }
%& \triangleq &
\left[\begin{array}{c c}
p & q \\
r & s
\end{array}\right].
\end{eqnarray*}

Our next goal is to use an orthogonal matrix to transform
$T_0 T$ to be upper-triangular.
%Since $Y$ is non-degenerate, 
$T_0$ is non-singular, therefore
$p$ and $r$ cannot both be zero. If $r=0$ then $T_0 T$ is
already upper-triangular. If $p=0$
%%% JYC: I think thse two cases can be combined.  as long as p^2 +r^2
%%% \not = 0
then the orthogonal matrix
$Q=\left[\begin{array}{c c}
0 & 1 \\
1 & 0
\end{array}\right]$
makes $QT_0T$ upper-triangular.
In general, if $p^2 +r^2 \not = 0$, then we can find
a (complex) orthogonal matrix $Q$ such that
$QT_0T$ is upper-triangular.
%%% JYC: use (c -s\\ s c). can show , use c= (e^{ix} + e^{-ix} )/2 etc
%%% can solve a quadratic eqn. etc.
It can be verified that $(QT_0T)^{\otimes 3} [1,1,0,0]^{\tt T}$,
where $QT_0T$ is upper-triangular,
has the form $[v, u, 0, 0]$ for some non-zero $u$ and $v$. We normalize it
to $[v,1,0,0]$.

A crucial observation is
 that for any orthogonal matrix $Q$, the LHS $[1,0,1]$
is {\it unchanged} under the holographic transformation $Q$:
$[1,0,1] (Q^{-1})^{\otimes 2} = [1,0,1]$.
This  gives us
\begin{displaymath}
\textrm{Holant}([1,0,1]|(T_0T)^{\otimes 3}[1,1,0,0]^{\tt T})  \equiv_{\rm T}
\textrm{Holant}([1,0,1]|[v,1,0,0]) \equiv_{\rm T}
\textrm{Holant}([v,1,0,0]),
\end{displaymath}
for some $v$. This shows the equivalence of the original instance
with $\textrm{Holant}([v,1,0,0])$.
By Lemma \ref{lemma-dichotomy-double-reoot}, $\textrm{Holant}
([v,1,0,0])$ is \#P-hard.

%Otherwise, both $p$ and $r$ are non-zero.
%If the above instance is tractable, it must be that $p^2+r^2=0$.
%Otherwise, we could find an orthogonal transformation,
%such that
%\begin{displaymath}
%\textrm{Holant}([1,0,1]|(T_0T)^{\otimes 3}[1,1,0,0]^T)  \equiv_{\rm T}
%\textrm{Holant}([v,1,0,0])
%\end{displaymath}
%for some $v$, and by Lemma \ref{lemma-dichotomy-double-reoot},
% $\textrm{Holant}([v,1,0,0])$ is \#P-hard.
%

Finally
$p^2+r^2=0$ implies that
$(a+ci)^2+(b+di)^2 =
(a^2+b^2)+2i(ac+bd)-(c^2+d^2) =0$.
This is exactly $y_2 - 2i y_1 - y_0 = 0$.
%(c^2+d^2)-2i(ac+bd)-(a^2+b^2) & = & 0.

\end{proof}

%% main theorem
\begin{theorem}\label{thm:main}
Let $\mathscr{F}$ be a set of complex symmetric signatures.
${\rm Holant}^c(\mathscr{F})$ is \#P-hard unless $\mathscr{F}$
satisfies one of the following conditions, in which case it is
tractable:
\begin{enumerate}
	\item ${\rm Holant}^*(\mathscr{F})$ is tractable (for which
we have an effective dichotomy---Theorem~\ref{thm-holant-star});
%~\cite{STOC09});
or
	\item There exists a $T \in \mathscr{T}$ such that
%%% JYC: I just called it \mathscr{T}
		$\mathscr{F} \subseteq T\mathscr{A}$, where
$
\mathscr{T} \triangleq \{T \mid
	[1,0,1]T^{\otimes 2}, [1,0]T, [0,1]T \in \mathscr{A}\}
$.
\end{enumerate}
\end{theorem}

\begin{proof}
First of all, if $\mathscr{F}$ is an exceptional
 case of Lemma \ref{lm-exist-ternary},
we know that  ${\rm Holant}^*(\mathscr{F})$ is tractable and we are done.
%Now we can assume that this is not the case,
%and as a result we can reduce the
%problem to ${\rm Holant}^*(\mathscr{F}\cup \{X\})$
%by Lemma \ref{lm-exist-ternary}, where $X=[x_0,x_1,x_2,x_3]$ is a
% non-degenerate symmetric ternary signature.
%%%
%%% JYC I don't see this "reduce the problem to"  shouldn't it be the other
%%% direction?
%%% also shouldn't it be Holant^c ???
Now we can assume  that we have a
non-degenerate symmetric ternary signature $X=[x_0,x_1,x_2,x_3]$
and the problem is ${\rm Holant}^c(\mathscr{F}\cup \{X\})$.

As discussed in Section \ref{sec-ternary}, there are three
categories for $X$ and we only need to consider the first two:
\begin{enumerate}
	\item $x_k=\alpha_1^{3-k}\alpha_2^k+\beta_1^{3-k}\beta_2^k$;
	\item $x_k=Ak\alpha^{k-1}+B\alpha^k$, where $A \neq 0$;
\end{enumerate}

\noindent
\emph{Case 1: $x_k=\alpha_1^{3-k}\alpha_2^k+\beta_1^{3-k}\beta_2^k$. }

In this case, $X=T^{\otimes 3}[1,0,0,1]^T$, where  $T=\left[
	\begin{array}{c c}
		\alpha_1 & \beta_1 \\
		\alpha_2 & \beta_2
	\end{array}
\right]$. So we have the following reduction chain,
\begin{eqnarray*}
{\rm Holant}^c(\mathscr{F}) & \equiv_{\rm T} &
 {\rm Holant}^c(\mathscr{F} \cup \{X\})
   \equiv_{\rm T}   {\rm Holant}(\mathscr{F} \cup \{X, [1,0],[0,1]\})   \\
&  \equiv_{\rm T} &
	{\rm Holant}(\{[1,0,1],[1,0],[0,1]\}|\mathscr{F} \cup \{X\}) \\
%~~\text{(unaries can be assumed to connect to non-unaries in $\mathscr{F} \cup \{X\}$)}
	& \equiv_{\rm T} &
	{\rm Holant}(\{[1,0,1]T^{\otimes 2},[1,0]T,[0,1]T\}|
[1,0,0,1] \cup  T^{-1}\mathscr{F} ).
\end{eqnarray*}

Since  $[1,0,1]T^{\otimes 2}$ is a non-degenerate binary signature,
we can apply Theorem \ref{thm-bipartite}.
The only thing we need to verify is that in the
%case $[1,0,1]T^{\otimes 2}= [\alpha_1^2+\alpha_2^0, \alpha_1 \beta_1 + \alpha_2
% \beta_2, \beta_1^2+\beta_2^0 ]=[0,\alpha_1 \beta_1 + \alpha_2 \beta_2,0]$,
case $[1,0,1]T^{\otimes 2}= [\alpha_1^2+\alpha_2^0, \alpha_1 \beta_1 + \alpha_2
\beta_2, \beta_1^2+\beta_2^2] = [0,\alpha_1 \beta_1 + \alpha_2 \beta_2,0]$,
%%% JYC pl check. i think the exponent 0 is wrong.
%%%
at least one of $[1,0]T=[\alpha_1, \beta_1]$ or
 $[0,1]T=[\alpha_2, \beta_2]$ has both entries non-zero.
 If not, we would have $\alpha_1 \beta_1=0$ and $\alpha_2 \beta_2=0$, which implies that
 $[1,0,1]T^{\otimes 2}= [0,\alpha_1 \beta_1 + \alpha_2 \beta_2,0]=[0,0,0]$, a contradiction.
 Therefore, by Theorem \ref{thm-bipartite}, we know that
the problem is \#P-hard unless
 $[1,0,1]T^{\otimes 2}  \cup T^{-1}\mathscr{F} \subseteq \mathscr{P}$
(note that unary $[1,0]T,[0,1]T$ are automatically in $\mathscr{P}$)
 or  $\{ [1,0,1]T^{\otimes 2} ,[1,0]T,[0,1]T \} \cup T^{-1}\mathscr{F} \subseteq \mathscr{A}$.
In the first case,  ${\rm Holant}^*(\mathscr{F})$ is tractable;
 in the second
case, this is equivalent to having  $T \in \mathscr{T}$ satisfying
		$\mathscr{F} \subseteq T\mathscr{A}$.

\noindent
\emph{Case 2: $x_k=Ak\alpha^{k-1}+B\alpha^k$, where $A \neq 0$. }

%%% JYC: I renamed it k, instead of i, since we will have alpha = i

In this case, if $\alpha \neq \pm i$, the problem is \#P-hard
by Lemma~\ref{lemma-dichotomy-double-reoot}
 and we are done. Now we consider
the case $\alpha = i$ (the case $\alpha = -i$ is similar).
%%%
%%%
Consider the following Equation
\begin{equation}\label{three-term-eqn}
z_{k+2} -2 i z_{k+1} -z_k=0.
\end{equation}
We note that $X = [x_0,x_1,x_2,x_3]$ satisfies this equation for
$k=0,1$.
\begin{comment}
If for all non-degenerate signatures
$Z = [z_0,z_1,\ldots, z_m]$ in $\mathscr{F}$
with arity $m \geq 2$, $Z$ satisfies
Equation (\ref{three-term-eqn}) for $k=0,1,\ldots, m-2$,
then, by Theorem~\ref{thm-holant-star}  (tractable case 2),
${\rm Holant}^*(\mathscr{F})$ is tractable, and we are done.
\end{comment}
If all non-degenerate signatures
$Z = [z_0,z_1,\ldots, z_m]$ in $\mathscr{F}$
with arity $m \geq 2$ satisfy the following

\noindent
{\bf Condition}: $Z$ satisfies
Equation (\ref{three-term-eqn}) for $k=0,1,\ldots, m-2$.

\noindent
then, by Theorem~\ref{thm-holant-star}  (tractable case 2),
${\rm Holant}^*(\mathscr{F})$ is tractable, and we are done.
So suppose this is not the case,
and $Z = [z_0,z_1,\ldots, z_m] \in \mathscr{F}$, for some $m \ge 2$,
is a non-degenerate signature that does not satisfy this Condition.
By Lemma~\ref{lemma-double-root}, if any non-degenerate
sub-signature $[z_k, z_{k+1}, z_{k+2}]$ does not satisfy
Equation (\ref{three-term-eqn}),
 then, together with $X$ which does satisfy (\ref{three-term-eqn}),
we know that
 the problem is \#P-hard and we are done.
So we assume every non-degenerate
sub-signature $[z_k, z_{k+1}, z_{k+2}]$ of $Z$ satisfies
(\ref{three-term-eqn}). In particular $m \ge 3$, and
there exists some binary sub-signature of $Z$ that is
degenerate and does not satisfy (\ref{three-term-eqn}).

\noindent
\emph{Subcase 1:} All binary sub-signatures of  $Z$
%$[z_0,z_2,\ldots, z_m]$
are degenerate (but $Z$ itself is non-degenerate).

We claim that $Z$ has the form $[z_0,0,\ldots,0, z_m]$, where $z_0 z_m
\not = 0$.
For a contradiction suppose $z_0 =0$, since $Z$ is non-degenerate,
there exists $k < m$ such that $z_k \not = 0$. Let $k$ be the
minimum such, then $0< k < m$ and $[z_{k-1}, z_k, z_{k+1}]$ is
non-degenerate.  So $z_0  \not = 0$ and similarly $z_m \not = 0$.
If there are any other $0 < k < m$ such that $z_k \not = 0$,
then let $k$ be the
minimum such. If $k=1$, then a simple induction shows that
$z_k = z_0 (z_1/z_0)^k$, for $0 \le k \le m$, and $Z$
is degenerate. If $k>1$, then $[z_{k-1}, z_k, z_{k+1}]$ is
non-degenerate, a contradiction.

 Next we claim that there exists a unary sub-signature
 $[x_k, x_{k+1}]$ of $X$  with both entries
non-zero. If $x_0 x_1 \neq 0$, then we set $k=0$ and the claim is
proved; if $x_0=0, x_1 \neq 0$, then we have $x_2= 2 i x_1+ x_0 \neq 0$;
 if  $x_0 \neq 0, x_1 = 0 $,
then we have $x_2= 2 i x_1+ x_0 \neq 0$ and  $x_3= 2 i x_2+ x_1 \neq 0$.
(Note that $x_0=x_1=0$ is impossible because  $A \neq 0$.)

 Connect this unary signature
 to $m-3$ dangling edges of $[z_0,0,\ldots,0, z_m]$, we
have a ternary signature
 $[a,0,0,b]$ where $ab\neq 0$.
We can use this as the non-degenerate
ternary signature $X$ and we have reduced this case to Case 1.

\noindent
\emph{Subcase 2:}
Some binary sub-signatures of $[z_0,z_1,\ldots, z_m]$ are non-degenerate
(some others are degenerate).

 Then we can find a ternary sub-signature
$[z_k, z_{k+1}, z_{k+2}, z_{k+3}]$ (or its reversal)
where $[z_k, z_{k+1}, z_{k+2}]$ is degenerate
and $[z_{k+1}, z_{k+2}, z_{k+3}]$ is non-degenerate
and thus satisfies $-z_{k+1}-2 i z_{k+2}+ z_{k+3}=0$.
%%% JYC : i have yet to check
%%% JYC see my example. i am not convinced.
%%% [0,0,c, 2ic]
If $\{z_{k}, z_{k+1}, z_{k+2}\}$ is a geometric sequence of a non-zero ratio $p$,
we could assume that $z_k=1$ after scaling. Then we have
\begin{displaymath}
\left[\begin{array}{c c}
	1 & 0 \\
	p & (p+2i p^2-p^3)^{1/3}
\end{array}\right]^{\otimes 3}[1,0,0,1]^{\tt T}=[z_k,z_{k+1},z_{k+2},z_{k+3}]^{\tt T}
\end{displaymath}
Therefore, the ternary sub-signature $[z_k, z_{k+1}, z_{k+2}, z_{k+3}]$ is in the first category
and we reduce the problem to Case 1.
Otherwise, it must be that $z_k=z_{k+1}=0$ and $z_{k+2} \neq 0$. This signature became
$[0,0,z_{k+2},z_{k+3}]$, which is equivalent to $[0,0,1,v]$ for some $v \in \mathbb{C}$,
and as we proved in Section 3, it is \#P-hard.
\end{proof}

%% file: propositions.tex
\section{Some Known Dichotomy Results}\label{Known-Dichotomy}
In this section, we review three dichotomy theorems from \cite{STOC09}.

\begin{theorem}\label{thm-holant-star}\cite{STOC09}
%%% JYC : I think we insist in this notion, this set $\mathscr{F}$ is after
%%% removal of all degenerate signatures? as I recall.
%%%  (as we talk about Holant^*)
Let $\mathscr{F}$ be a set of symmetric non-degenerate
 signatures over $\mathbb{C}$.
Then $\rm{Holant}^*(\mathscr{F})$ is computable in polynomial time
in the following three cases. In all other cases,
$\rm{Holant}^*(\mathscr{F})$ is \#P-hard.
\begin{enumerate}
	\item Every signature in $\mathscr{F}$ is of arity no more than two;
	\item There exist two constants $a$ and $b$ (not both zero, depending
		only on $\mathscr{F}$), such that for every signature
		$[x_0,x_1,\ldots,x_n] \in \mathscr{F}$ one of the two
 conditions is
		satisfied: (1) for every $k=0,1,\ldots,n-2$, we have
		$ax_{k}+bx_{k+1}-ax_{k+2}=0$; (2) $n=2$ and the signature
		$[x_0,x_1,x_2]$ is of form $[2a\lambda,b\lambda,-2a\lambda]$.
	\item For every signature $[x_0,x_1,\ldots,x_n] \in \mathscr{F}$, one
		of the two conditions is satisfied: (1) For every $k=0,1,\ldots,n-2$,
		we have $x_{k}+x_{k+2}=0$; (2) $n=2$ and the signature $[x_0,x_1,x_2]$
		is of form $[\lambda,0,\lambda]$.
\end{enumerate}
\end{theorem}

\begin{theorem}\label{thm:holnat-c}\cite{STOC09}
Let $\mathscr{F}$ be a set of \emph{real} symmetric signatures, and let
$\mathscr{F}_1, \mathscr{F}_2$ and $\mathscr{F}_3$
 be three families of signatures
defined as
\begin{eqnarray*}
	\mathscr{F}_1 & = &
		\{\lambda([1,0]^{\otimes k}+i^r[0,1]^{\otimes k}) |
			\lambda \in \mathbb{C}, k=1,2, \ldots, r=0,1,2,3\}; \\
	\mathscr{F}_2 & = &
		\{\lambda([1,1]^{\otimes k}+i^r[1,-1]^{\otimes k}) |
			\lambda \in \mathbb{C}, k=1,2, \ldots, r=0,1,2,3\}; \\
	\mathscr{F}_3 & = &
		\{\lambda([1,i]^{\otimes k}+i^r[1,-i]^{\otimes k}) |
			\lambda \in \mathbb{C}, k=1,2, \ldots, r=0,1,2,3\}.
\end{eqnarray*}
Then ${\rm Holant}^c (\mathscr{F})$ is computable in polynomial time if
(1) After removing unary signatures from $\mathscr{F}$,
it falls in one of the three cases of Theorem~\ref{thm-holant-star}
(this implies  ${\rm Holant}^* (\mathscr{F})$ is computable in polynomial time)
 or
(2) (Without removing any unary signature)
$\mathscr{F} \subseteq \mathscr{F}_1 \cup \mathscr{F}_2 \cup \mathscr{F}_3$.
 Otherwise, ${\rm Holant}^c (\mathscr{F})$ is \#P-hard.
\end{theorem}

\begin{definition}
A $k$-ary function $f(x_1,\ldots,x_k)$ is affine if it
has the form
\begin{displaymath}
\chi_{AX=0} \cdot \sqrt{-1}^{\sum_{j=1}^{n}\langle \alpha_j,X\rangle}
\end{displaymath}
where $X=(x_1,x_2,\ldots,x_k,1)$, $A$ is matrix over $\mathbb{F}_2$,
$\alpha_j$ is a vector over $\mathbb{F}_2$,
  and $\chi$ is a 0-1 indicator function
such that $\chi_{AX=0}$ is 1 iff $AX=0$. Note that the inner
product $\langle \alpha_j, X\rangle$
 is calculated over $\mathbb{F}_2$, while the
summation $\sum_{j=1}^{n}$ on the exponent of $i = \sqrt{-1}$
 is evaluated as a sum mod 4 of 0-1 terms.
%over $\mathbb{F}_4$. 
We use $\mathscr{A}$
to denote the set of all affine functions.

We use $\mathscr{P}$
to denote the set of functions which can be expressed as a
product of unary functions, binary equality functions ($[1,0,1]$)
and binary disequality functions ($[0,1,0]$).
\end{definition}

\begin{theorem}\label{thm:dichotomy}\cite{STOC09}
Suppose $\mathscr{F}$ is a class of functions mapping Boolean inputs to
complex numbers. If $\mathscr{F} \subseteq \mathscr{A}$ or $\mathscr{F}
\subseteq \mathscr{P}$, then \#CSP($\mathscr{F}$) is computable in
polynomial time.
Otherwise, \#CSP($\mathscr{F}$) is {\rm \#P}-hard.
\end{theorem}

As we mentioned in \cite{STOC09}, the class $\mathscr{A}$ is a natural generalization of
the symmetric signatures family $\mathscr{F}_1 \cup \mathscr{F}_2 \cup \mathscr{F}_3$. 
It is easy to show that the set of symmetric signatures in  $\mathscr{A}$ is exactly
$\mathscr{F}_1 \cup \mathscr{F}_2 \cup \mathscr{F}_3$.

% We will also show that the tractable cases 2 and
%3 in Theorem \ref{thm-holant-star} for Holant$^*$ can be unified after a holographic reduction and is closely
%related the family $\mathscr{P}$. Based on these observations, we
%can restate these dichotomies in a different form. And these insights are important for the
%dichotomies in this paper.

\section{Some Useful Reductions}\label{useful-reductions}

In this section, we list some useful simple
reductions: reduction between Holant and \#CSP,
reduction between bipartite and non-bipartite settings, and holographic reduction.

\begin{proposition}
$\#{\rm CSP}(\mathscr{F}) \equiv_{\rm T}
	{\rm Holant}\left(\mathscr{F} \cup
		%\bigcup_{j=1}^{m}\{=_j\}\right) 
\bigcup_{j\ge 1} \{=_j\}\right)
%%% JYC: why stop in m?  what's m?
 \equiv_{\rm T} {\rm Holant}(\mathscr{F} \cup
		\{=_3\}) $.
\end{proposition}
This says that \#CSP is the same as Holant problems 
with {\sc Equality} functions given for free.

\begin{proposition}\label{prop2}
  $ {\rm Holant}(\mathscr{F})  \equiv_{\rm T}  {\rm Holant}([1,0,1]| \mathscr{F}) $.
\end{proposition}
That is,  we can transform every edge to a path of length 2
with the new vertex given  $(=_2) = [1,0,1]$.

%%%%%%%

\begin{proposition}
  $ {\rm Holant}(\mathscr{G}_1 \cup [1,0,1]  | \mathscr{G}_2 \cup  [1,0,1]  )  \equiv_{\rm T}  {\rm Holant}(\mathscr{G}_1 \cup \mathscr{G}_2) $.
\end{proposition}
Binary {\sc Equality} functions on both sides allow the transfer of
signatures.

\begin{proposition}\label{prop4}
For any $T \in {\bf GL}_2({\mathbb C})$,
  $ {\rm Holant}(\mathscr{G}_1  | \mathscr{G}_2  )  \equiv_{\rm T}  {\rm Holant}(\mathscr{G}_1 T | T^{-1} \mathscr{G}_2) $.
\end{proposition}
This is a restatement of Valiant's Holant Theorem. 

%%% JYC I changed above F to G, to avoid reusing the fixed usage of F_1
%%% F_2 F_3 in our 3 families.
%%%
\begin{proposition}
Let  $T$ be an orthogonal transformation ($T T^{\tt T}=I$). Then
${\rm Holant}(\mathscr{F})  \equiv_{\rm T}  {\rm Holant}(T \mathscr{F})$.
\end{proposition}
This follows from the invariance of $(=_2) = [1,0,1]$
under an orthogonal transformation, and Props.~\ref{prop2},  \& \ref{prop4}.

%% file: JYC-list-of-T.tex
\section{List of Matrices in $\mathscr{T}$}
In this section, we explicitly list all the matrices in the family
$$
\mathscr{T} \triangleq \{T \mid
	[1,0,1]T^{\otimes 2}, [1,0]T, [0,1]T \in \mathscr{A}\},
$$
which is defined and used in the statement of Theorem \ref{thm:main}.
The following condition is given in Theorem~\ref{thm:main}:

%\begin{quote}\label{condition-T}
\begin{equation}\label{condition-T}
\mbox{There exists a $T \in \mathscr{T}$ such that
                $\mathscr{F} \subseteq T\mathscr{A}$.}
\end{equation}
%\endquote}
Together with the condition in Theorem~\ref{thm-holant-star}, 
this gives an effective tractability condition which is both necessary
and sufficient
for ${\rm Holant}^c$ problems, by Theorem~\ref{thm:main}. 

As noted before, the set of symmetric signatures in  $\mathscr{A}$ is exactly
$\mathscr{F}_1 \cup \mathscr{F}_2 \cup \mathscr{F}_3$.
We note that $[1,0,1]T^{\otimes 2}$, $[1,0]T$ and $[0,1]T$ are all
symmetric, as is the requirement $T^{-1} \mathscr{F} \subseteq \mathscr{A}$
in Theorem~\ref{thm:main}. Thus we can replace $\mathscr{A}$
by $\mathscr{F}_1 \cup \mathscr{F}_2 \cup \mathscr{F}_3$
in the expression $\mathscr{T}$ above.
 
It is obvious that the family $\mathscr{T}$ is closed under a scalar
multiplication. Thus we list them up to a scalar multiple.
%After scaling, symmetric binary signatures in $\mathcal{F}$ are:
%%% JYC: check this symbol
After a scalar multiple, symmetric binary signatures in 
$\mathscr{F}_1 \cup \mathscr{F}_2 \cup \mathscr{F}_3$ are precisely
\[[1, 0, \pm 1], [1, 0, \pm i], [1, \pm 1, -1], [1, \pm i, 1], 
[0,1,0].\]
%We note that this set is invariant under reversal, up to a   
%scalar multiple.
Also the unary signatures in
$\mathscr{F}_1 \cup \mathscr{F}_2 \cup \mathscr{F}_3$, up to a 
scalar multiple,
are 
\[[1, \pm 1], [1,\pm i], [1,0], [0,1].\]

Before we enumerate all the possibilities, 
we make two observations to simplify this process:
\begin{enumerate}
\item If we exchange the two columns of $T$,
the signature $[1,0,1]T^{\otimes 2}$  becomes its reversal,
%remains unchanged,
%%% JYC:
%%%% if you do columns exchange, $[1,0,1]T^{\otimes 2}$ IS changed.
%%% eg try (1, 2 ; 3, 4)
%%%
%%% if you exchange two rows of T , it is unchagbed...
%%%This is because
%\[ [1,0,1]T^{\otimes 2} = ( (1, 0)^{\otimes 2} + (0, 1)^{\otimes 2} )
%T^{\otimes 2} = ( (1, 0)T)^{\otimes 2} + ((0, 1) T)^{\otimes 2}.\]
%%% i am not sure you want col or row. but looks ike you meant col.
%%%
		and the two numbers in $[1,0]T$ are
 interchanged. Similarly the
two numbers in $[0,1]T$ are interchanged as well.
 The effect of exchanging the two columns of $T$
is the same as replacing $T$ 
by $T \left[
        \begin{array}{c c}
                0 & 1 \\
                1 & 0
        \end{array}
\right]$. On the other hand,
the holographic transformation 
$\left[
        \begin{array}{c c}                
		0 & 1 \\
                1 & 0        
	\end{array}\right]
\mathscr{A}$ amounts exchanging input
values 0 and 1 for functions in $\mathscr{A}$,
and this operation keeps $\mathscr{A}$ invariant.
%%%
%%% JYC: 12-10-09: i am rewriting everything here below.
%%%
 %Therefore, we only need to consider
%		$[1,0]T \in \{[1,\pm 1], [1, 0], [1, i]\}$,
%up to a scalar factor,
%in order to determine whether there exists a $T \in \mathscr{T}$ such that
%                $\mathscr{F} \subseteq T\mathscr{A}$,
%as in Theorem~\ref{thm:main}.
%%% JYC. eg if there was another T' making [1,0]T' =[1, -i]
%%% then use another T, exchanging two column, can get to [-i, 1] equiv to
%%% [1, i]. the same T also keeps [1,0,1]T in A. and makes F in T A.
%%%
%%% but for later, when we "normalize" for [1 0 1]T
%%% we still need to keep [1,0]T = [0,1] case around.  I think. (JYC)
\item If we multiply $\left[\begin{array}{c c}1 & 0 \\0 & -1\end{array}\right]$ on the
		right side of $T$, then $[y_0,y_1,y_2] \triangleq [1,0,1]T^{\otimes 2}$ becomes
		$[y_0,-y_1,y_2]$. 
This operation also preserves both the set of binary and
the set of unary signatures,
respectively,
listed for $\mathscr{F}_1 \cup \mathscr{F}_2 \cup \mathscr{F}_3$,
up to  a scalar factor.
%The same invariance holds for the set of unary signatures
%listed for $\mathscr{F}_1 \cup \mathscr{F}_2 \cup \mathscr{F}_3$.
%%% this is to multiply -1 to the second entry.
On the other hand,
the effect of the holographic transformation
$\left[
        \begin{array}{c c}
                1 & 0 \\
                0 & -1   
        \end{array}\right]
\mathscr{A}$ is to transform an original
function $f$ to $f \cdot (-1)^{\sum_i x_i}$,
and thus it is in $\mathscr{A}$ iff the original $f \in \mathscr{A}$.
%Furthermore this operation does not affect $[1,0]T$ from item 1.
%%% JYC: actually i have a problem with this.
%%%  once you have used item 1 to make $[1,0]T$ as listed.
%%% can you still make this change of T by right multiply (1 0 ; 0, -1)?
%%% wouldn't that affect $[1,0]T$?  eg wouldn't i have to think
%%% about [1,0]T = (1 -i) from the case in item 1. [1,0]T = (1 i) ???
%%%
%So we only need to consider $T$'s such that
%		$[1,0,1]T^{\otimes 2} \in \{[1,0,\pm 1], [1,0,\pm i],
%[1,1,-1], [1,i,1],[0,1,0]\}$,
%up to a scalar factor.

Similarly, we can
 multiply $\left[\begin{array}{c c}1 & 0 \\ 0 & i\end{array}\right]$
on the right side of $T$. 
Of course the invariance under 
$\left[\begin{array}{c c}1 & 0 \\ 0 & i\end{array}\right]$
implies that of $\left[\begin{array}{c c}1 & 0 \\ 0 & -1\end{array}\right]
= \left[\begin{array}{c c}1 & 0 \\ 0 & i\end{array}\right]^2$.
%Then we only need to investigate
%$[1,0,1]T^{\otimes 2}
% \in \{[1,0,1], [1,0,i], [1,i,1],[0,1,0]\}$,
%up to a scalar factor.
%
%
\end{enumerate}

What has been shown is that  Condition~(\ref{condition-T})
is invariant under the right action on $\mathscr{T}$ by the group
generated by $\left[
        \begin{array}{c c}
                0 & 1 \\
                1 & 0
        \end{array}\right]$
and 
 $\left[\begin{array}{c c}1 & 0 \\ 0 & i\end{array}\right]$.
By $\left[\begin{array}{c c}1 & 0 \\ 0 & i\end{array}\right]$,
we may  consider only those $T$'s such that
$[1,0,1]T^{\otimes 2}
 \in \{[1,0,1], [1,0,i], [1,i,1],[0,1,0]\}$,
up to a scalar factor.
If we further normalize by the reversal action
$\left[
        \begin{array}{c c}
                0 & 1 \\
                1 & 0
        \end{array}\right]$
we may  consider only those $T$'s such that
$[1,0]T \in \{[1,\pm 1], [1, i], [1, 0]\}$,
up to a scalar factor.
However this reversal action is only partially closed
for $\{[1,0,1], [1,0,i], [1,i,1],[0,1,0]\}$,
with the exception $[1,0,i]$ which is changed to $[1,0,-i]$.
Thus we may have two extra cases to consider:
$[1,0,1]T^{\otimes 2} = [1,0,i]$ and, $[1,0]T = [0,1]$ or $[1,-i]$.
But for these two cases if we apply 
first $\left[\begin{array}{c c}1 & 0 \\ 0 & i\end{array}\right]$, 
followed by 
$\left[
        \begin{array}{c c}
                0 & 1 \\
                1 & 0
        \end{array}\right]$,
we obtain $[1,0,1]T^{\otimes 2} = [1,0,i]$ and,
$[1,0]T = [1,0]$ or $[1,1]$ respectively.
 Hence we can eliminate these two cases.
%%% JYC
%%% for [1,0,i] | [0,1]: after ( 1 0; 0 i) it is [1,0,-i]|[0,i]
%%% which is the same as [1,0,-i]|[0,1] by scalar
%%% then reversal gets [1,0,i] | [1,0]
%%% for [1,0,i] | [1,-i]: after ( 1 0; 0 i) it is [1,0,-i]|[1,1]
%%% then reversal gets [1,0,i] | [1,1]

To summarize,
to enumerate all $T$ satisfying
Condition~(\ref{condition-T})
we only need to consider
\[
[1,0,1]T^{\otimes 2}
 \in \{[1,0,1], [1,0,i], [1,i,1],[0,1,0]\}~~~\mbox{  and  }~~~
[1,0]T \in \{[1,\pm 1], [1, i], [1, 0]\},\]
up to a scalar factor.
In the following, we denote by $\alpha=(1+i)/\sqrt{2}=\sqrt{i}$.

\vspace{.2in}

%%% JYC
%%% I knew I said someting that we can remove the second one in each
%%% pair of matrices T in each case, where the secon is obtained from
%%% the first by * (-1) on second row.
%%% but now I am not certain we can omit those.
%%% it is true that for the condition {script T} this multiply -1
%%% to second row does not affect [1,0,1]T and [1,0]T and [0,1]T in {script A}
%%% but I am not sure it does not affect the condition 
%%% given F \subseteq T {script A}.
%%% if we replace T by this * (-1) on second row of T, ie (1 0; 0, -1) * T
%%% it is same as preprocessing the given F by replacing each f \in F
%%% with (-1)^{\sum_i x_i} f.  and then check against the old T
%%% to see if we get all f in F are in T {script A}.
%%%

%%% so unless there is an argument, we should restore the second half 
%%% of the T's listed.

%%%%%%%%%JYC 12-09-09 i am restoring the second half of each pair in T.%%%%%%

%%% I don't see why you need to both \lambda and \gamma below.
%%% one is enough. i think.

\input{actual-list-of-T.tex}

%% file: actual-list-of-T.tex
If $[1,0,1]T^{\otimes 2}=\gamma[1,0,1]$, $[1,0]T=\lambda[1,1]$, we have 
\begin{displaymath}
T=
\left[
	\begin{array}{c c}
		1 & 1 \\
		1 & -1
	\end{array}
\right],
\left[
	\begin{array}{c c}
		1 & 1 \\
		-1 & 1
	\end{array}
\right].
\end{displaymath}

If $[1,0,1]T^{\otimes 2}=\gamma[1,0,1]$, $[1,0]T=\lambda[1,-1]$, we have
\begin{displaymath}
T=
\left[
	\begin{array}{c c}
		1 & -1 \\
		1 & 1
	\end{array}
\right],
\left[
	\begin{array}{c c}
		1 & -1 \\
		-1 & -1
	\end{array}
\right].
\end{displaymath}

%%% JYC:
%%% I don't see you needed to put \pm i there in [1, \pm i]. changed.
%%%
For $[1,0,1]T^{\otimes 2}=\gamma[1,0,1]$, $[1,0]T=\lambda[1, i]$, there is no solution.

If $[1,0,1]T^{\otimes 2}=\gamma[1,0,1]$, $[1,0]T=\lambda[1,0]$, we have
\begin{displaymath}
T=
\left[
	\begin{array}{c c}	
		1 & 0 \\
		0 & \pm 1
	\end{array}
\right].
\end{displaymath}

If $[1,0,1]T^{\otimes 2}=\gamma[1,0,i]$, $[1,0]T=\lambda[1,1]$, we have
\begin{displaymath}
T=
\left[
	\begin{array}{c c}
		1 & 1 \\
		\alpha^3 & \alpha
	\end{array}
\right],
\left[
	\begin{array}{c c}
		1 & 1 \\
		-\alpha^3 & -\alpha
	\end{array}
\right].
\end{displaymath}

If $[1,0,1]T^{\otimes 2}=\gamma[1,0,i]$, $[1,0]T=\lambda[1,-1]$, we have
\begin{displaymath}
T=
\left[
	\begin{array}{c c}
		1 & -1 \\
		\alpha^3 & -\alpha
	\end{array}
\right],
\left[
	\begin{array}{c c}
		1 & -1 \\
		-\alpha^3 & \alpha
	\end{array}
\right].
\end{displaymath}

If $[1,0,1]T^{\otimes 2}=\gamma[1,0,i]$, $[1,0]T=\lambda[1,i]$, we have
\begin{displaymath}
T=
\left[
	\begin{array}{c c}
		1 & i \\
		\alpha & -\alpha
	\end{array}
\right],
\left[
	\begin{array}{c c}
		1 & i \\
		-\alpha & \alpha
	\end{array}
\right].
\end{displaymath}

If $[1,0,1]T^{\otimes 2}=\gamma[1,0,i]$, $[1,0]T=\lambda[1,0]$, we have
\begin{displaymath}
T=
\left[
	\begin{array}{c c}
		1 & 0 \\
		0 & \alpha
	\end{array}
\right],
\left[
	\begin{array}{c c}
		1 & 0 \\
		0 & -\alpha
	\end{array}
\right].
\end{displaymath}

If $[1,0,1]T^{\otimes 2}=\gamma[1,i,1]$, $[1,0]T=\lambda[1,1]$, we have
\begin{displaymath}
T=
\left[\begin{array}{c c}
	1 & 1 \\
	-\alpha^3 & \alpha^3
\end{array}\right],
\left[\begin{array}{c c}
	1 & 1 \\
	\alpha^3 & -\alpha^3
\end{array}\right].
\end{displaymath}

If $[1,0,1]T^{\otimes 2}=\gamma[1,i,1]$, $[1,0]T=\lambda[1,-1]$, we have
\begin{displaymath}
T=\left[\begin{array}{c c}
	1 & -1 \\
	\alpha & \alpha
\end{array}\right],
\left[\begin{array}{c c}
	1 & -1 \\
	-\alpha & -\alpha
\end{array}\right].
\end{displaymath}

If $[1,0,1]T^{\otimes 2}=\gamma[1,i,1]$, $[1,0]T=\lambda[1,i]$, we have
\begin{displaymath}
T=\left[\begin{array}{c c}
	1 & i \\
	0 & \sqrt{2}
\end{array}\right],
\left[\begin{array}{c c}
	1 & i \\
	0 & -\sqrt{2}
\end{array}\right].
\end{displaymath}

If $[1,0,1]T^{\otimes 2}=\gamma[1,i,1]$, $[1,0]T=\lambda[1,0]$, we have
\begin{displaymath}
T=
\left[\begin{array}{c c}
	\sqrt{2} & 0 \\
	i & 1
\end{array}\right],
\left[\begin{array}{c c}
	\sqrt{2} & 0 \\
	-i & 1
\end{array}\right].
\end{displaymath}

If $[1,0,1]T^{\otimes 2}=\gamma[0,1,0]$, $[1,0]T=\lambda[1,1]$, we have
\begin{displaymath}
T=
\left[\begin{array}{c c}
	1 & 1 \\
	i & -i
\end{array}\right],
\left[\begin{array}{c c}
	1 & 1 \\
	-i & i
\end{array}\right].
\end{displaymath}

If $[1,0,1]T^{\otimes 2}=\gamma[0,1,0]$, $[1,0]T=\lambda[1,-1]$, we have
\begin{displaymath}
T=
\left[\begin{array}{c c}
	1 & -1 \\
	-i & -i
\end{array}\right],
\left[\begin{array}{c c}
	1 & -1 \\
	i & i
\end{array}\right].
\end{displaymath}

If $[1,0,1]T^{\otimes 2}=\gamma[0,1,0]$, $[1,0]T=\lambda[1,i]$, we have
\begin{displaymath}
T=
\left[\begin{array}{c c}
	1 & i \\
	-i & -1
\end{array}\right],
\left[\begin{array}{c c}
	1 & i \\
	i & 1
\end{array}\right].
\end{displaymath}

For $[1,0,1]T^{\otimes 2}=\gamma[0,1,0]$, $[1,0]T=\lambda[1,0]$, there is no solution.

%% file: orthogonal.tex
\section{An Orthogonal Transformation}\label{Orthogonal}

In this Section 
we give the detail of an {\it orthogonal} holographic
transformation used in the proof of Lemma~\ref{lemma-dichotomy-double-reoot}.

We are given
 $x_k=A k \alpha^{k-1}+B \alpha^k$, where $A \not = 0$,
and $\alpha \neq \pm i$. Let
$S = {\begin{bmatrix} 1 & \frac{B-1}{3} \\
\alpha &   A+ \frac{B-1}{3} \alpha \end{bmatrix}}$, then the
signature $[x_0, x_1, x_2, x_3]$ can be expressed as
 \[(x_0, x_1, x_1, x_2, x_1, x_2, x_2, x_3)^{\tt T}
=  S^{\otimes 3}
   (1, 1, 1, 0, 1, 0, 0, 0)^{\tt T}.\]
This identity can be verified by observing that
\[ (1, 1, 1, 0, 1, 0, 0, 0)^{\tt T} =
 \begin{bmatrix} 1 \\ 0 \end{bmatrix}^{\otimes 3}
+ \begin{bmatrix} 1 \\ 0 \end{bmatrix} \otimes
  \begin{bmatrix} 1 \\ 0 \end{bmatrix} \otimes
  \begin{bmatrix} 0 \\ 1 \end{bmatrix}
+ \begin{bmatrix} 1 \\ 0 \end{bmatrix} \otimes
  \begin{bmatrix} 0 \\ 1 \end{bmatrix} \otimes
  \begin{bmatrix} 1 \\ 0 \end{bmatrix}
+ \begin{bmatrix} 0 \\ 1 \end{bmatrix} \otimes
  \begin{bmatrix} 1 \\ 0 \end{bmatrix} \otimes
  \begin{bmatrix} 1 \\ 0 \end{bmatrix},\]
and we apply $S^{\otimes 3}$ using properties of tensor product,
$S^{\otimes 3} \begin{bmatrix} 1 \\ 0 \end{bmatrix}^{\otimes 3}
= (S \begin{bmatrix} 1 \\ 0 \end{bmatrix})^{\otimes 3}$, etc.

Let $T = \frac{1}{\sqrt{1 +
\alpha^2}}
{ \begin{bmatrix} 1 & \alpha \\
\alpha & -1 \end{bmatrix}}$,
then $T = T^{\tt T} = T^{-1} \in  {\bf O}_2(\mathbb{C})$ is  orthogonal,
and $R = T S = { \begin{bmatrix} u & w \\
0 & v \end{bmatrix}}$ is upper triangular, where $u = \sqrt{1 +
\alpha^2}$. As $\det R = \det T \det S = (-1) A \not = 0$,
we have $uv \not = 0$.
It follows that
\begin{eqnarray*}
&& T^{\otimes 3}  (x_0, x_1, x_1, x_2, x_1, x_2, x_2, x_3)^{\tt T}\\
&=& (TS)^{\otimes 3} (1, 1, 1, 0, 1, 0, 0, 0)^{\tt T}\\
&=& R^{\otimes 3} (1, 1, 1, 0, 1, 0, 0, 0)^{\tt T} \\
&=& R^{\otimes 3} \left\{
\begin{bmatrix} 1 \\ 0 \end{bmatrix}^{\otimes 3}
+ \begin{bmatrix} 1 \\ 0 \end{bmatrix} \otimes
  \begin{bmatrix} 1 \\ 0 \end{bmatrix} \otimes
  \begin{bmatrix} 0 \\ 1 \end{bmatrix}
+ \begin{bmatrix} 1 \\ 0 \end{bmatrix} \otimes
  \begin{bmatrix} 0 \\ 1 \end{bmatrix} \otimes
  \begin{bmatrix} 1 \\ 0 \end{bmatrix}
+ \begin{bmatrix} 0 \\ 1 \end{bmatrix} \otimes
  \begin{bmatrix} 1 \\ 0 \end{bmatrix} \otimes
  \begin{bmatrix} 1 \\ 0 \end{bmatrix} \right\}\\
&=&
\begin{bmatrix} u \\ 0 \end{bmatrix}^{\otimes 3}
+ \begin{bmatrix} u \\ 0 \end{bmatrix} \otimes
  \begin{bmatrix} u \\ 0 \end{bmatrix} \otimes
  \begin{bmatrix} w \\ v \end{bmatrix} 
+ \begin{bmatrix} u \\ 0 \end{bmatrix} \otimes
  \begin{bmatrix} w \\ v \end{bmatrix} \otimes
  \begin{bmatrix} u \\ 0 \end{bmatrix} 
+ \begin{bmatrix} w \\ v \end{bmatrix} \otimes
  \begin{bmatrix} u \\ 0 \end{bmatrix} \otimes
  \begin{bmatrix} u \\ 0 \end{bmatrix}
\end{eqnarray*}
This can be written as a  symmetric signature form $[u^3 + 3u^2 w,
u^2v, 0, 0]$. Note that the entry $u^2v \not = 0$,
which we can normalize to 1, after a scalar multiplication.
This gives us the  form $[z, 1, 0, 0]$
for some $z \in \mathbb{C}$.

%% file: ccc.bbl
\begin{thebibliography}{}

\end{thebibliography}


\begin{thebibliography}{10}
\bibitem{Bulatov06}
Andrei~A. Bulatov.
\newblock A dichotomy theorem for constraint satisfaction problems on a
  3-element set.
\newblock {\em J. ACM}, 53(1):66--120, 2006.

\bibitem{Bulatov08}
Andrei~A. Bulatov.
\newblock The complexity of the counting constraint satisfaction problem.
\newblock In {\em ICALP (1)}, volume 5125 of {\em Lecture Notes in Computer Science},
  pages 646--661. Springer, 2008.

\bibitem{BulatovD03}
Andrei~A. Bulatov and V\'{\i}ctor Dalmau.
\newblock Towards a dichotomy theorem for the counting constraint satisfaction
  problem.
\newblock {\em Inf. Comput.}, 205(5):651--678, 2007.

\bibitem{BulatovG04}
Andrei~A. Bulatov and Martin Grohe.
\newblock The complexity of partition functions.
\newblock In {\em ICALP}, volume 3142 of {\em Lecture Notes in
  Computer Science}, pages 294--306. Springer, 2004.

\bibitem{BulatovG05}
Andrei~A. Bulatov and Martin Grohe.
\newblock The complexity of partition functions.
\newblock {\em Theor. Comput. Sci.}, 348(2-3):148--186, 2005.

\bibitem{Homomorphisms}
Jin-Yi Cai, Xi~Chen, and Pinyan Lu.
\newblock Graph homomorphisms with complex values: A dichotomy theorem.
\newblock {\em manuscript}, 2009.


\bibitem{STOC07}
Jin-Yi Cai and Pinyan Lu.
\newblock Holographic algorithms: from art to science.
\newblock In {\em STOC '07: Proceedings of the thirty-ninth annual ACM
  symposium on Theory of computing}, pages 401--410, 2007.

\bibitem{STACS07}
Jin-Yi Cai and Pinyan Lu.
\newblock On symmetric signatures in holographic algorithms.
\newblock In {\em STACS}, volume 4393
  of {\em Lecture Notes in Computer Science}, pages 429--440. Springer, 2007.

\bibitem{FOCS08}
Jin-Yi Cai, Pinyan Lu, and Mingji Xia.
\newblock Holographic algorithms by fibonacci gates and holographic reductions
  for hardness.
\newblock In {\em FOCS '08: Proceedings of the 49th Annual IEEE Symposium on
  Foundations of Computer Science}, pages 644--653,  2008.

\bibitem{TAMC}
Jin-Yi Cai, Pinyan Lu, and Mingji Xia.
\newblock A Computational Proof of Complexity of Some Restricted Counting
               Problems
\newblock TAMC 2009: 138-149.

 \bibitem{STOC09}
Jin-Yi Cai, Pinyan Lu, and Mingji Xia.
\newblock Holant Problems and Counting CSP.
\newblock STOC 2009: 715-724.



\bibitem{CSPBook}
N.~Creignou, S.~Khanna, and M.~Sudan.
\newblock {\em Complexity classifications of boolean constraint satisfaction
  problems}.
\newblock SIAM Monographs on Discrete Mathematics and Applications, 2001.

\bibitem{DagumL92}
P.~Dagum and M.~Luby.
\newblock Approximating the permanent of graphs with large factors.
\newblock {\em Theor. Comput. Sci.}, 102:283--305, 1992.

\bibitem{dodson}
C.~T.~J. Dodson and T.~Poston.
\newblock {\em Tensor Geometry}.
\newblock Graduate Texts in Mathematics 130. Springer-Verlag, New York, 1991.

\bibitem{weightedCSP}
Martin~E. Dyer, Leslie~Ann Goldberg, and Mark Jerrum.
\newblock The complexity of weighted boolean \#csp.
\newblock {\em CoRR}, abs/0704.3683, 2007.

\bibitem{DyerGP06}
Martin~E. Dyer, Leslie~Ann Goldberg, and Mike Paterson.
\newblock On counting homomorphisms to directed acyclic graphs.
\newblock In {\em ICALP (1)}, volume 4051 of {\em Lecture Notes in
  Computer Science}, pages 38--49. Springer, 2006.

\bibitem{acyclic}
Martin~E. Dyer, Leslie~Ann Goldberg, and Mike Paterson.
\newblock On counting homomorphisms to directed acyclic graphs.
\newblock {\em J. ACM}, 54(6), 2007.

\bibitem{DyerG00}
Martin~E. Dyer and Catherine~S. Greenhill.
\newblock The complexity of counting graph homomorphisms (extended abstract).
\newblock In {\em SODA}, pages 246--255, 2000.

\bibitem{DyerG04}
Martin~E. Dyer and Catherine~S. Greenhill.
\newblock Corrigendum: The complexity of counting graph homomorphisms.
\newblock {\em Random Struct. Algorithms}, 25(3):346--352, 2004.

\bibitem{FederV98}
Tom{\'a}s Feder and Moshe~Y. Vardi.
\newblock The computational structure of monotone monadic snp and constraint
  satisfaction: A study through datalog and group theory.
\newblock {\em SIAM J. Comput.}, 28(1):57--104, 1998.

\bibitem{freedman-l-s}
M.~Freedman, L.~Lov\'{a}sz, and A.~Schrijver.
\newblock Reflection positivity, rank connectivity, and homomorphism of graphs.
\newblock {\em J. AMS}, 20:37--51, 2007.

\bibitem{GJGT}
Leslie~Ann Goldberg, Martin Grohe, Mark Jerrum, and Marc Thurley.
\newblock A complexity dichotomy for partition functions with mixed signs.
\newblock {\em CoRR}, abs/0804.1932, 2008.

\bibitem{Hell}
P.~Hell and J.~Ne\v{s}et\v{r}il.
\newblock On the complexity of h-coloring.
\newblock {\em Journal of Combinatorial Theory, Series B}, 48(1):92--110, 1990.

\bibitem{Kasteleyn1961}
P.~W. Kasteleyn.
\newblock The statistics of dimers on a lattice.
\newblock {\em Physica}, 27:1209--1225, 1961.

\bibitem{Kasteleyn1967}
P.~W. Kasteleyn.
\newblock Graph theory and crystal physics.
\newblock In {\em Graph Theory and Theoretical Physics},
  pages 43--110. Academic Press, London, 1967.

\bibitem{mike}
Michael Kowalczyk.
\newblock Classification of a Class of Counting Problems Using Holographic   Reductions.
\newblock In \emph{Proceedings of COCOON '09}, pages 472-485, 2009.

\bibitem{cai}
Michael Kowalczyk and Jin-Yi Cai.
\newblock Holant Problems for Regular Graphs with Complex Edge Functions.
\newblock To appear in STACS 2010.

\bibitem{TF1961}
H.~N.~V. Temperley and M.~E. Fisher.
\newblock Dimer problem in statistical mechanics - an exact result.
\newblock {\em Philosophical Magazine}, 6:1061--1063, 1961.

\bibitem{Vadhan01}
Salil~P. Vadhan.
\newblock The complexity of counting in sparse, regular, and planar graphs.
\newblock {\em SIAM J. Comput.}, 31(2):398--427, 2001.

\bibitem{Valiant79b}
Leslie~G. Valiant.
\newblock The complexity of enumeration and reliability problems.
\newblock {\em SIAM J. Comput.}, 8(3):410--421, 1979.

\bibitem{HA_FOCS}
Leslie~G. Valiant.
\newblock Holographic algorithms (extended abstract).
\newblock In {\em FOCS '04: Proceedings of the 45th Annual IEEE Symposium on
  Foundations of Computer Science}, pages 306--315,2004.
SIAM J. Comput. 37(5): 1565-1594 (2008).

\bibitem{AA_FOCS}
Leslie~G. Valiant.
\newblock Accidental algorthims.
\newblock In {\em FOCS '06: Proceedings of the 47th Annual IEEE Symposium on
  Foundations of Computer Science}, pages 509--517,  2006.

\bibitem{XiaZZ07}
Mingji Xia, Peng Zhang, and Wenbo Zhao.
\newblock Computational complexity of counting problems on 3-regular planar
  graphs.
\newblock {\em Theor. Comput. Sci.}, 384(1):111--125, 2007.

\end{thebibliography}
